\documentclass{article}

 \PassOptionsToPackage{numbers}{natbib}
\usepackage[dandb,preprint]{neurips_2025}

\usepackage{times}
\usepackage{latexsym}
\usepackage{xcolor}         
\usepackage{wrapfig,lipsum,booktabs}

\usepackage[T1]{fontenc}

\usepackage[utf8]{inputenc}

\usepackage{microtype}

\usepackage{inconsolata}

\usepackage{graphicx}
\usepackage{subcaption}
\usepackage[hang,flushmargin]{footmisc}
\usepackage{enumitem}
\usepackage{color, colortbl}
\usepackage{xspace}
\usepackage{booktabs}
\usepackage{multirow}
\definecolor{citecolor}{RGB}{34,139,34}
\usepackage{mdframed}
\usepackage[pagebackref=true,breaklinks=true,colorlinks,
citecolor=citecolor,bookmarks=false]{hyperref}
\usepackage{cleveref}

\newcommand\ignore[1]{}

\newcommand{\searcharena}{Search Arena\xspace}
\newcommand{\gpt}{GPT\textsubscript{4.1}\xspace}
\newcommand{\ma}{model$_A$\xspace}
\newcommand{\mb}{model$_B$\xspace}
\newcommand{\an}{Auto\-Nuggetizer\xspace}

\title{Chatbot Arena Meets Nuggets: Towards Explanations and Diagnostics in the Evaluation of LLM Responses}

\author{Sahel Sharifymoghaddam\thanks{Equal Contribution},  Shivani Upadhyay\footnotemark[1], Nandan Thakur\footnotemark[1], \\\textbf{Ronak Pradeep, Jimmy Lin} \\[1ex]
David R.\ Cheriton School of Computer Science,\\
 University of Waterloo, Canada \\[1ex]
\texttt{\{sahel.sharifymoghaddam, sjupadhyay, nandan.thakur,} \\
\texttt{ rpradeep, jimmylin@\}uwaterloo.ca}}

\begin{document}
\maketitle

\begin{abstract}
Battles, or side-by-side comparisons in so-called arenas that elicit human preferences, have emerged as a popular approach for assessing the output quality of LLMs.
Recently, this idea has been extended to retrieval-augmented generation (RAG) systems.
While undoubtedly representing an advance in evaluation, battles have at least two drawbacks, particularly in the context of complex information-seeking queries:\ they are neither explanatory nor diagnostic.
Recently, the nugget evaluation methodology has emerged as a promising approach to evaluate the quality of RAG answers.
Nuggets decompose long-form LLM-generated answers into atomic facts, highlighting important pieces of information necessary in a ``good'' response.
In this work, we apply our \an framework to analyze data from roughly 7K Search Arena battles provided by LMArena in a fully automatic manner.
Our results show a significant correlation between nugget scores and human preferences, showcasing promise in our approach to explainable and diagnostic system evaluations. All the code necessary to reproduce results in our work is available in \url{https://github.com/castorini/lmsys\_nuggetize}.
\end{abstract}

\section{Introduction}

The notion of ``battles'', or side-by-side comparisons of responses from large language models (LLMs), has become a popular method for evaluating their quality~\cite{mt-bench, chiang2024chatbot}.
In the ``arena'' setup, users are shown two LLM outputs and asked to indicate which one they prefer.
This approach was popularized by LMSYS through MT-Bench~\cite{mt-bench} and later expanded into the Chatbot Arena~\cite{chiang2024chatbot}.
The popularity of these arenas has made them a key marketing tool when launching new LLMs from companies such as Google, OpenAI, and Meta, who regularly tout leaderboard rankings on Chatbot Arena in model releases.
Inspired by this increased popularity, arena-based evaluations have been extended to a variety of domains, including AI agents~\cite{agent-arena}, vision and image generation~\cite{lu2024wildvision, jiang2024genai}, and multilingual generation~\cite{thakur-mirage-bench}.

Most recently, battles were extended to search-augmented LLMs in the \searcharena~\cite{searcharena2025}.
Unlike the original setup, which focused on ``closed-book'' generation of responses by LLMs directly, \searcharena evaluates systems that apply retrieval-augmented generation (RAG) to retrieve relevant source documents, which are then used by LLMs to generate long-form answers with citations~\cite{ragnarok,han-etal-2024-rag}.
While such side-by-side comparisons enable the evaluation of search-augmented LLM-based systems at scale, we see them having at least two drawbacks:\ they are neither explanatory nor diagnostic, especially in scenarios where determining the better answer is not straightforward. 
It would be desirable for an evaluation to (at least attempt to) explain {\it why} a user might have preferred one response over another.
Furthermore, we argue that evaluations should be diagnostic in providing actionable guidance on how to improve systems. 

We hypothesize that the recently introduced nugget evaluation methodology~\cite{autonuggetizer, greatnuggetrecall} can be adapted to potentially address these two drawbacks for complex information-seeking queries.
The basic idea is to assess answer quality in terms of the recall of information nuggets, or atomic facts, that should be present in good responses.
In our previous work, we have shown that this can be accomplished in a fully automatic manner using LLMs.

\begin{figure*}[t]
\centering
\begin{center}
    \includegraphics[width=\textwidth]{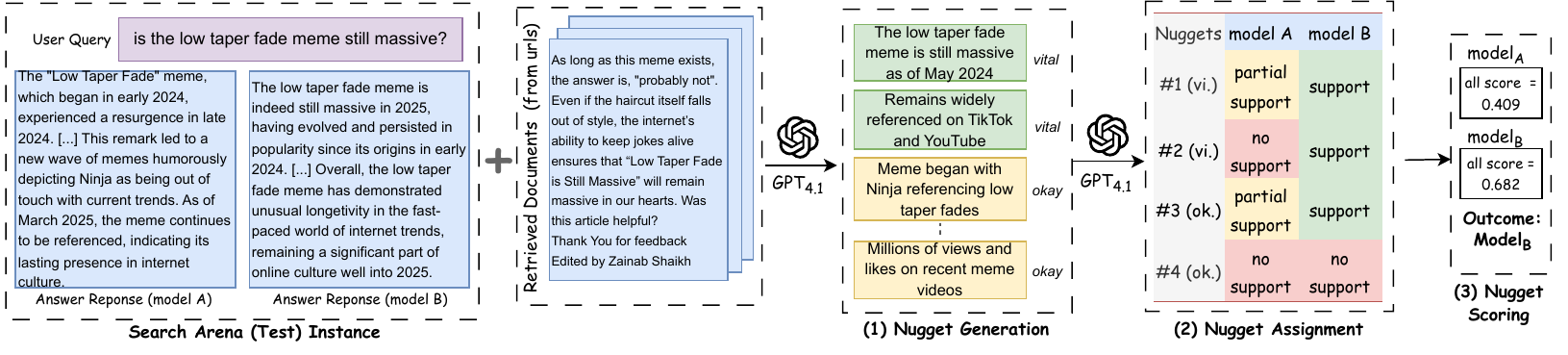}
    \caption{An end-to-end example from Search Arena illustrating both nugget generation and assignment. First, \gpt generates nuggets based on the query, retrieved chunks from URL contents, and the responses from both models. Each nugget is then labeled with an importance level—either ``vital'' or ``okay''. Next, \gpt evaluates whether each model supports each nugget, assigning one of three labels: ``support'', ``partial support'', or ``no support''. Finally, these support judgments are scored and aggregated to determine the overall outcome (the model with the higher score is preferred).}
    \label{fig:overview}
\end{center}
\vspace*{-\baselineskip}
\end{figure*}

In this paper, we adapt the \an~\cite{autonuggetizer} implementation of the nugget evaluation methodology to analyze recently released public data from \searcharena in a fully automatic manner (see~\autoref{fig:overview}).
We find that human preferences correlate well with the distribution of nugget scores, conditioned on these preferences, which can be plotted as density functions.
Further analyses reveal that these distributions differ significantly from one another, providing strong evidence for the explanatory power of nugget scores in capturing human preferences.
From here, nugget score differences provide actionable guidance to improve RAG systems.

\section{Related Work}

\paragraph{Nugget-based evaluation.}
Originally introduced in the TREC QA Track in 2003~\citep{voorhees-2003-evaluating-answers, Voorhees_TREC2003}, the nugget-based evaluation methodology focuses on identifying essential atomic facts---called nuggets---that are relevant to a given question.
This methodology was later extended to tasks like summarization and broader conceptions of question answering~\citep{nenkova-passonneau-2004-evaluating, lin-demner-fushman-2006-will, Dang_Lin_ACL2007, Lin_Zhang_SIGIR2007}, and researchers have explored automation to improve its scalability~\citep{Lin_Demner-Fushman_HLT-EMNLP2005, Lin_Demner-Fushman_IR2006, nuggetir}.

The recent emergence of large language models (LLMs) has enabled automated, reliable nugget-based evaluation~\citep{autonuggetizer, alaofi2024generative, greatnuggetrecall, thakur-freshstack, 2025conversationalgold}. 
Several RAG evaluation frameworks---such as FactScore~\citep{min-etal-2023-factscore}, RUBRIC~\citep{farzi:2024}, and others~\citep{arabzadeh2024comparison, machine_gen}---incorporate the nugget concept, although most of these proposed approaches are either not validated or primarily validated on traditional ad hoc retrieval, and hence their applicability to long-form answers is unclear.
We refer readers to Pradeep et al. \cite{greatnuggetrecall} for a more detailed discussion of related work.
Here, we use the \an framework from Pradeep et al. \cite{autonuggetizer}, and apply it to the side-by-side comparisons of LLM responses within the Search Arena.

\paragraph{Related arena benchmarks.}
The \searcharena, introduced by LMArena~\citep{searcharena2025}, is a recently introduced (April 14, 2025) and popular benchmark evaluating LLMs with access to a search tool. 
Other notable efforts include the MTEB Arena~\cite{mtebarena}, which extends the Massive Text Embedding Benchmark (MTEB) framework~\cite{mteb} to head-to-head evaluation across embedding models, and Ragnar\"{o}k~\citep{ragnarok}, which released a new variant of the MS MARCO collection (V2.1) and offers a framework for evaluation in the TREC 2024 RAG Track.

\section{Experimental Design}\label{sec:experimental-design}

\paragraph{\searcharena overview.}

The \searcharena, introduced by LMArena in Miroyan et al. \cite{searcharena2025}, is a crowd-sourced evaluation platform for search-augmented LLM systems, evaluated in terms of side-by-side comparisons that solicit human preferences~\cite{chiang2024chatbot}.
The V1 version of the publicly released dataset\footnote{\url{https://huggingface.co/datasets/lmarena-ai/search-arena-v1-7k}} contains in total 7K samples for which two RAG-focused systems denoted as \ma and \mb (e.g., Gemini-2.5-Pro-Grounding vs.~Perplexity-Sonar-Reasoning-Pro) battle each other. 
For each battle, a human assessor judges with one of four responses, whether \ma (or \mb) is the winner or a tie where both are responses are good (or bad).
Each battle also includes search result URLs used by the search-augmented LLM systems for response generation when available---approximately 6.7K out of 7K battles.
In total, the dataset contains around 80K unique URLs.

The \searcharena dataset comprises both single-turn and multi-turn battles.
In this work, we focus exclusively on single-turn battles, where the system provides a single response back to the user, evaluating a total of 5,103 instances.
We exclude multi-turn battles from our experiments because human votes at the overall battle level do not reliably reflect turn-level (per-query) preferences, which is what \an evaluates.

\searcharena also contains battles for several non-English languages, e.g., Chinese or Russian. 
Non-English languages collectively account for less than 40\% of the dataset, with English comprising the remaining majority.
Detailed statistics for single-turn battles used in this work are presented in~\Cref{app:dataset_stats_sec}. 
Queries in \searcharena vary widely, ranging from long code snippets to prompts that demand complex reasoning or exhibit ambiguity and vagueness.
We show a few examples of queries from the Search Arena dataset in~\Cref{app:query_classification}.

\paragraph{Corpus generation.}
To evaluate LLM responses in the absence of ground-truth answers, we use the URLs provided in the dataset, collected from each system response as relevant sources of information. 
We begin by constructing a corpus from the 47K unique URLs associated with single-turn battles. 
This process involves downloading the contents of each URL, extracting the main textual content, and segmenting the text into chunks of ten sentences with an overlap of two sentences, using spaCy’s~\footnote{\url{https://spacy.io/}} \texttt{xx\_sent\_ud\_sm} model.

Once the corpus is prepared, we encode the chunks and the query prompts utilizing the \texttt{BAAI/bge-m3} model. 
We then perform dense retrieval by computing cosine similarity between the chunk and query embeddings, implemented via Pyserini’s FAISS indexing and search~\cite{pyserini}. 
With this setting, we retrieve the top 50 most relevant chunks for each query.
Notably, both the chunking and encoding models support multilingual corpora, ensuring robust language coverage across the dataset.

\begin{figure}[t]
        \centering
        \includegraphics[width=0.6\columnwidth]{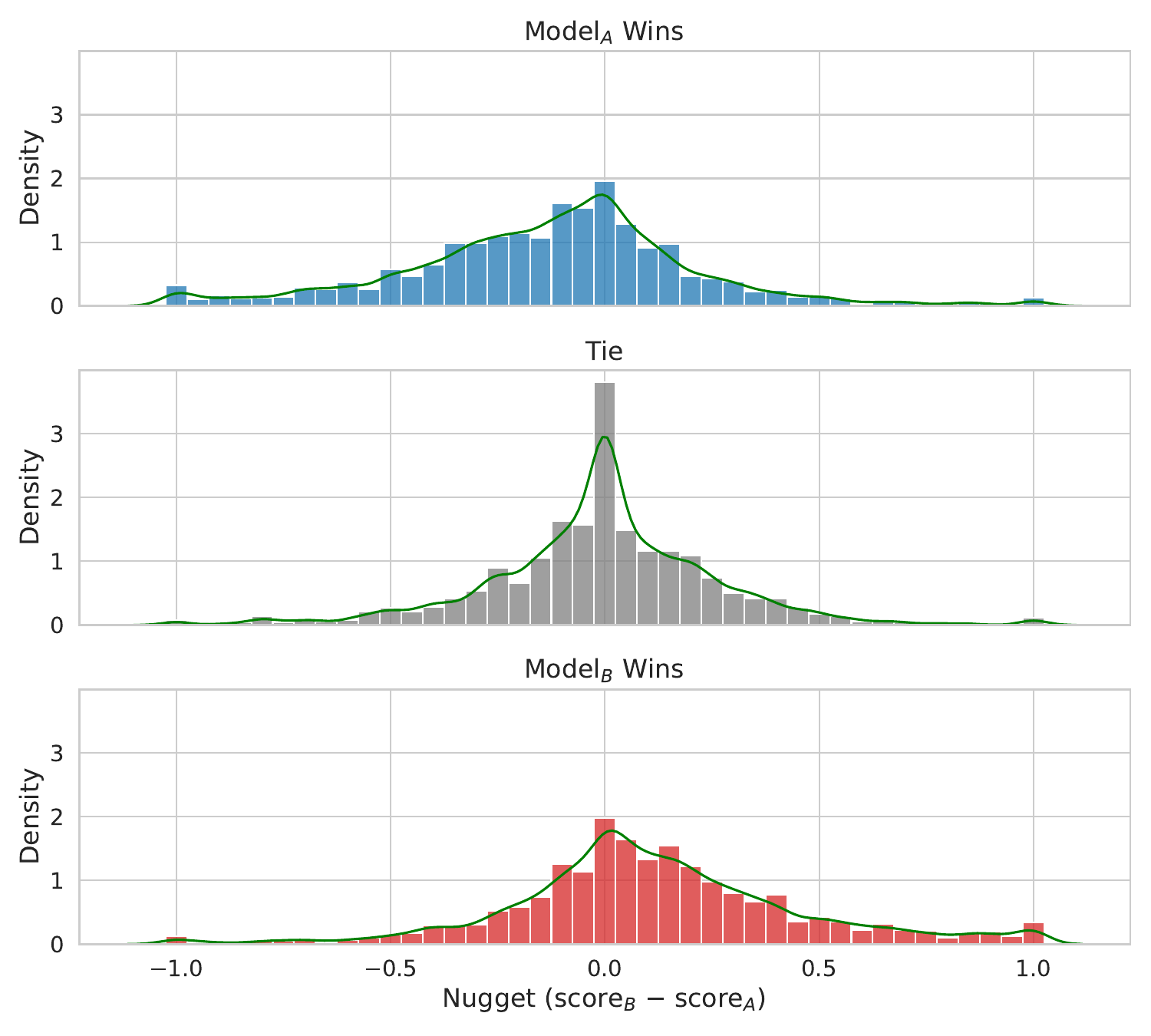}

    \caption{Empirical probability density function (PDF) of nugget score differences ($\textrm{score}_B$ $-$ $\textrm{score}_A$) grouped by human preference category: \ma wins, tie, or \mb wins. A Kernel Density Estimation (KDE) with a bandwidth of 0.5 is fitted separately for each group.}
    \label{fig:density}
\end{figure}

\paragraph{Nugget evaluation.}
Nugget generation creates atomic facts that highlight the essential information required in a RAG answer, and assignment categorizes their support level for the model response. 
Following Pradeep et al.~\cite{autonuggetizer}, we use the \an tool in the \texttt{nuggetizer} code repository\footnote{\url{https://github.com/castorini/nuggetizer}} to generate and assign information nuggets to model responses.
As shown in~\autoref{fig:overview}, there are two steps in nugget generation and assignment:

\begin{enumerate}[leftmargin=0.5cm]

\item \textbf{Nugget generation:} For each prompt extracted from the dataset, we construct a request to \an that includes the query (i.e., the prompt itself), along with relevant chunks retrieved from our created corpus, ordered by relevance.
In addition to these chunks, we include the responses from each model, inserted in random order to mitigate the positional bias.
We include model responses for two key reasons. First, approximately 5\% of the battles do not contain any URLs. Second, even when URLs are present, about 16\% of them yield 100 bytes or less of content after scraping--typically due to issues such as cookie or JavaScript requirements, invalid or expired links, geo-blocking, and similar obstacles.
These cases make the LLM responses a valuable fallback source of information for nugget generation.

The \an tool then processes the request and identifies nuggets that are relevant to the query from the retrieved chunks and the provided LLM responses. Furthermore, each nugget is assigned an importance label:\ ``vital'' or ``okay'', reflecting its relevance or significance to the input query.
Following previous guidelines, ``vital'' nuggets must be present in a ``good'' answer, while ``okay'' nuggets are good to have, but are not necessary.
Nugget importance labeling is run in a separate pass, independent of the actual responses. 

\item \textbf{Nugget assignment:} Once nuggets and their importance labels are generated (from the previous step), we use \an to assign them to model responses, determining whether each nugget is supported in the answer. This step categorizes each nugget into ``supported'', ``partially supported'', or ``not supported''. 
Among the four combinations of evaluation methods---``vital'' vs. ``all'' (vital + okay) and ``strict'' vs. ``non-strict'' (full + partial support)---we adopt the ``All Score'' metric, which achieves the highest recall by counting nuggets of all importance and support levels.
We find that the ``Strict Vital'' metric, which is the primary metric used in the TREC 2024 RAG Track~\cite{greatnuggetrecall}, is too strict for our use case, particularly when only a small number of nuggets are available.
\end{enumerate}

\noindent We emphasize that while the \an framework supports different degrees of manual intervention, here we are running the entire evaluation pipeline end-to-end automatically.

\section{Experimental Results}\label{sec:experimental-results}

All experiments in this paper are conducted using \gpt, the latest language model from OpenAI with a knowledge cutoff of June 2024, as the underlying model used by the \an via Microsoft Azure.
Out of the 5,103 single-turn battles in the \searcharena dataset, 5 were excluded from our analysis due to issues such as Azure content filtering, invalid output formats, or other nugget generation failures. 
On average, each single-turn battle full evaluation (comprising both nuggetization and assignment) requires approximately 2--3 seconds when executed using the Azure OpenAI API.
The mean number of nuggets generated per battle is $\sim$12.5.

\autoref{fig:density} presents our main results, the probability densities of nugget score differences ($\textrm{score}_B$ $-$ $\textrm{score}_A$) conditioned on the human preference judgment (i.e., the battle outcomes).
On the top, we show the distribution when model$_A$ wins; on the bottom, we show the distribution when \mb wins; and in the middle, ties.
Battles where the output of both models is considered to be equally bad are excluded from the distributions.

These results appear to support our hypothesis that nugget scores correlate with human preferences.
In the case where \ma wins (top row), the distribution skews to the left (negative values), indicating that \ma typically gets higher nugget scores than \mb.
Conversely, when \mb wins (bottom row), the distribution skews to the right (positive values), suggesting that \mb generally obtains a higher nugget score.
When the human indicates a tie (middle row), the distribution peaks around zero, as expected, indicating similar nugget scores between models.

\begin{figure*}[t]
        \includegraphics[width=\textwidth]{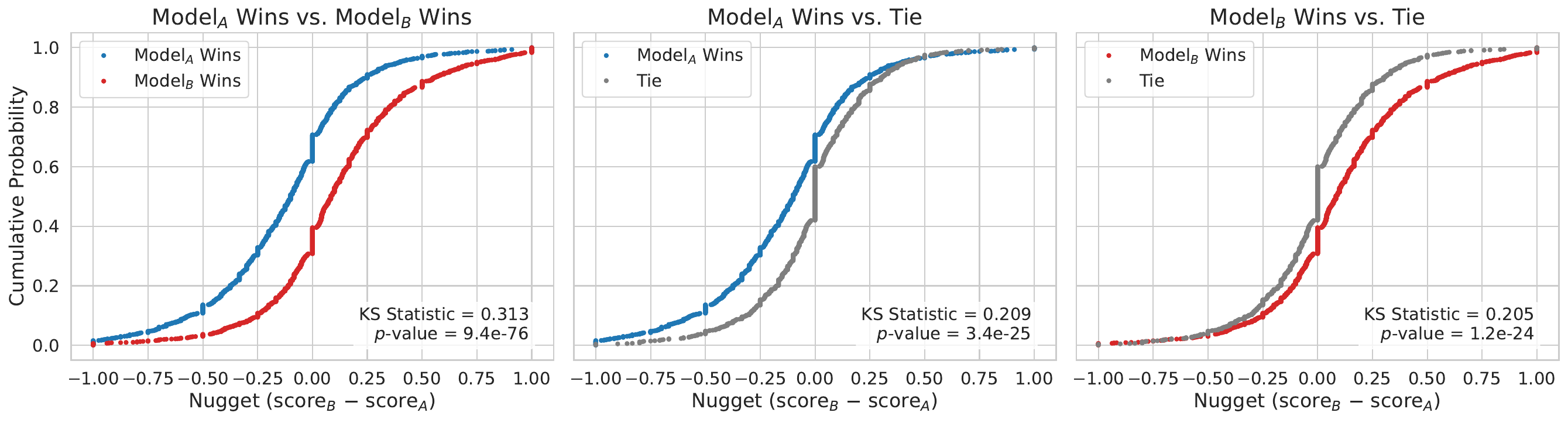}
        \caption{Empirical cumulative distribution functions (CDFs) comparing nugget score differences ($\textrm{score}_B$ $-$ $\textrm{score}_A$) across human vote categories. Each subplot shows a Kolmogorov-Smirnov (K-S) test between two groups: (left) \ma wins vs. \mb wins, (center) \ma wins vs. tie, and (right) \mb wins vs. tie. The K-S statistic and corresponding $p$-value are annotated in each plot, quantifying the distributional differences between groups.}
    \label{fig:ks_test}
\end{figure*}

To analyze the statistical differences among these three conditional distributions, we performed pairwise Kolmogorov-Smirnov (K-S) tests.
As shown in~\autoref{fig:ks_test}, the K-S statistic values range from 0.205 to 0.313, with $p$-values of $1.2\mathrm{e}^{-24}$ or lower, indicating that all three distributions differ significantly from one another (i.e., we have high confidence that these samples were drawn from different distributions).
These findings validate our hypothesis that nugget score differences align with human preferences, reinforcing the potential of nugget-based metrics as reliable evaluators of model quality in head-to-head evaluations.

\begin{wrapfigure}{r}{0.4\columnwidth}
    \centering
    \includegraphics[width=0.4\columnwidth]{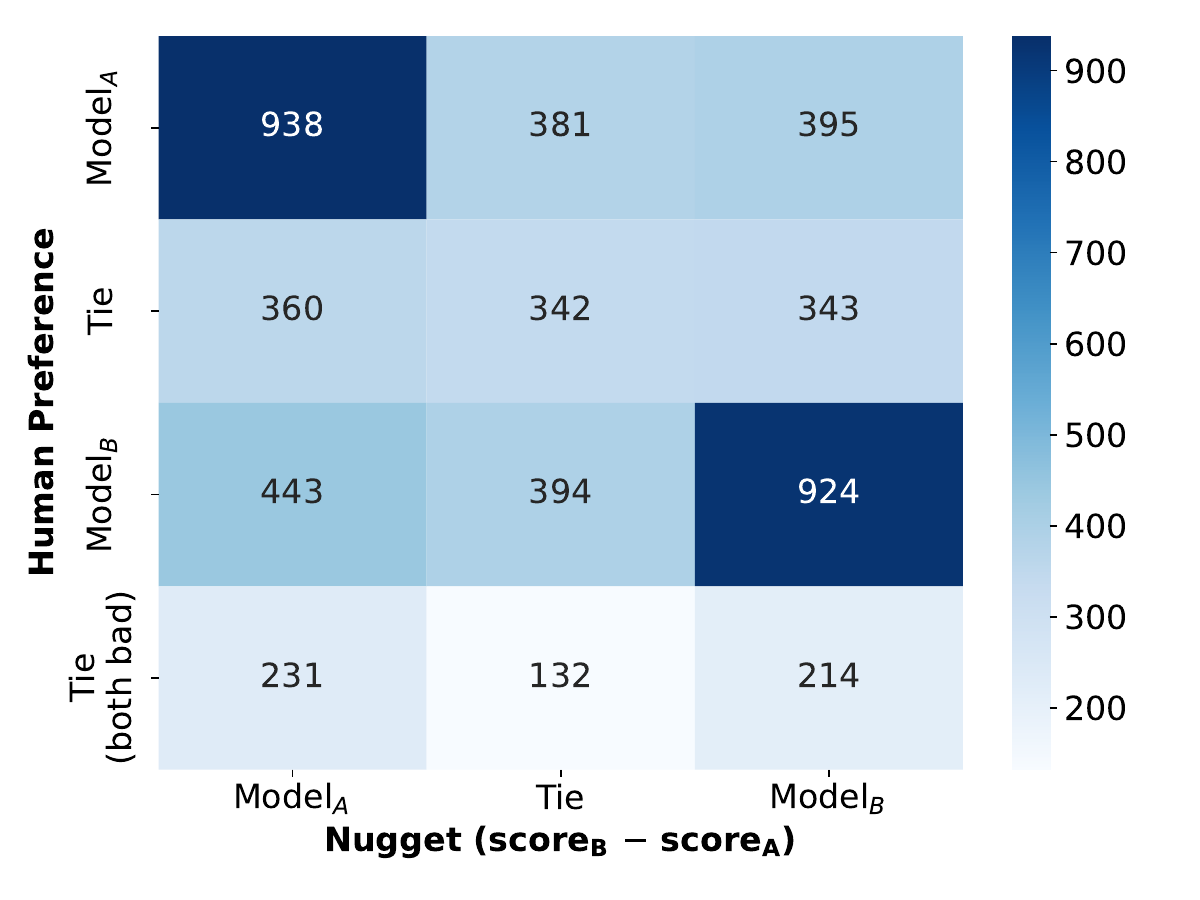}
    \caption{Confusion matrix comparing human and nugget preferences. 
    A threshold of 0.07 is applied to treat nugget preference scores as a tie.}
    \label{fig:confusion-matrix-all-score}
\end{wrapfigure}

\autoref{fig:confusion-matrix-all-score} presents a confusion matrix that visually compares the distribution of human preferences (rows) in \searcharena against ``nugget preferences'' (columns).
For ``nugget preference'', we use a threshold of 0.07, meaning that when the nugget score difference between the two model outputs falls within $\pm~0.07$, the comparison is considered a tie.
This threshold was selected by sweeping values between 0.05 and 0.15 in increments of 0.01.
A threshold of 0.07 results in nugget preferences that most closely reflect an equal distribution of \ma wins, \mb wins, and ties when the human preference is a tie (second row in~\autoref{fig:confusion-matrix-all-score}).
The diagonal cells in this confusion matrix reveal the instances where nugget preferences align with human preferences.
Conversely, off-diagonal cells illustrate the types and frequencies of disagreements between the human and nugget scores. 

In particular, the nugget-based evaluation prefers \ma in 938 out of 1,714 (54.7\%) of the battles where \ma wins the battle (first row in~\autoref{fig:confusion-matrix-all-score}).
Similarly, \mb is preferred in 924 out of 1761 (52.5\%) battles where it wins the battle (third row in~\autoref{fig:confusion-matrix-all-score}). 
We investigate the anti-diagonal corners where nugget and human preferences disagree--in the first two studies below. 
Then, we examine how the availability of URL contents affects nugget generation. 
Finally, we compare LLM-as-a-judge preference agreement with human preferences as an alternative to relying on the nugget-based preferences.

\subsection{Query Classification Analysis}

In this analysis, we use query classification to better understand the cases where nugget preferences and human preferences are not aligned.
When the nugget scores and the human prefer opposite sides of a battle, we refer to this situation as a ``preference inversion'', or simply inversion.

We suspect that inversions might vary across different types of queries.
To investigate, we followed Rosset et al. \cite{researchy} but used the newer \gpt to rate each query on a scale of 0--10 across eight different categories. 
Then, we classify each query into its maximum scoring category or categories (allowing for ties).
To further strengthen the category signals, we exclude queries with a maximum score less than seven from this classification.
Raw distributions of the query ratings per category and sample queries from each class are available in~\Cref{app:query_classification}.

\begin{figure*}[t!]
    \centering
    \begin{subfigure}[t]{0.24\textwidth}
        \centering
        \includegraphics[width=\textwidth]{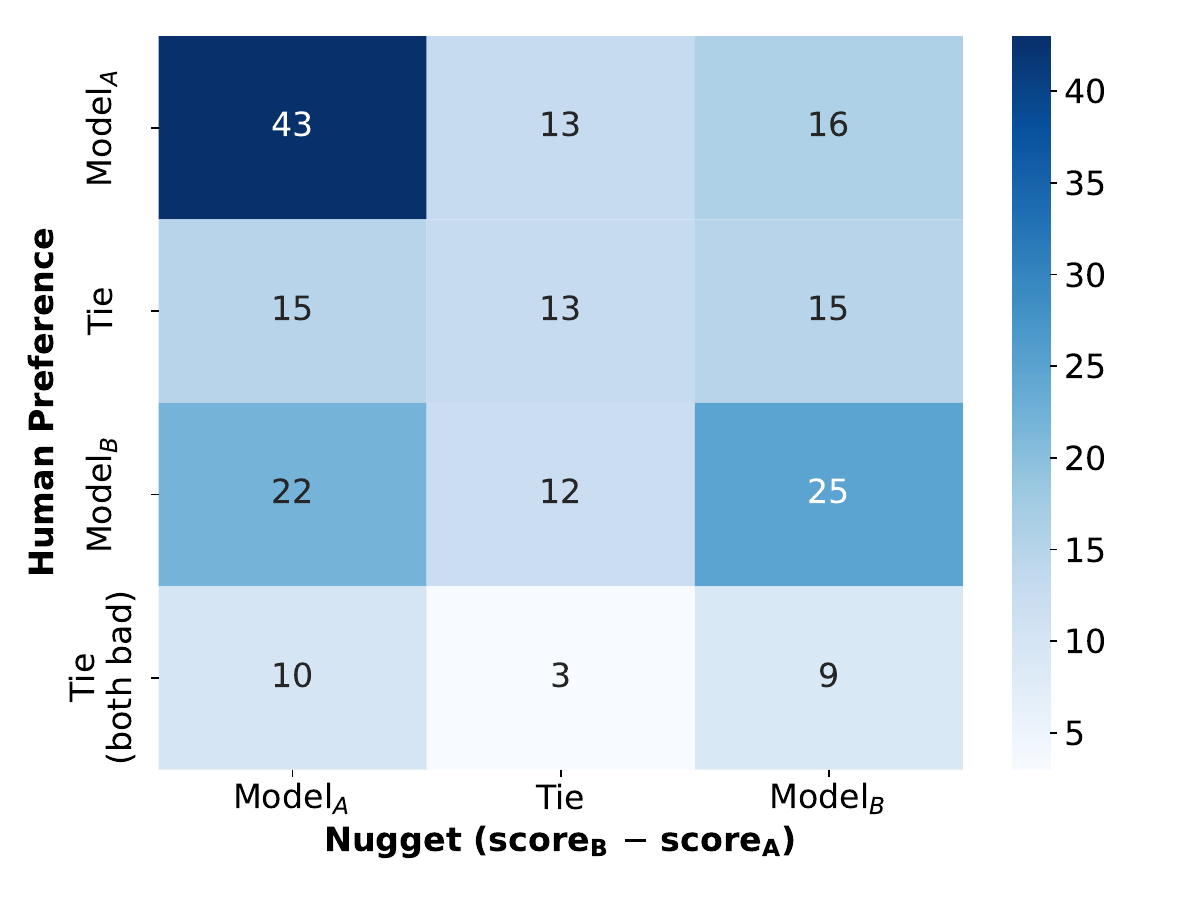}
        \caption{Ambiguous}
    \end{subfigure}
    \hfill
    \begin{subfigure}[t]{0.24\textwidth}
        \centering
        \includegraphics[width=\textwidth]{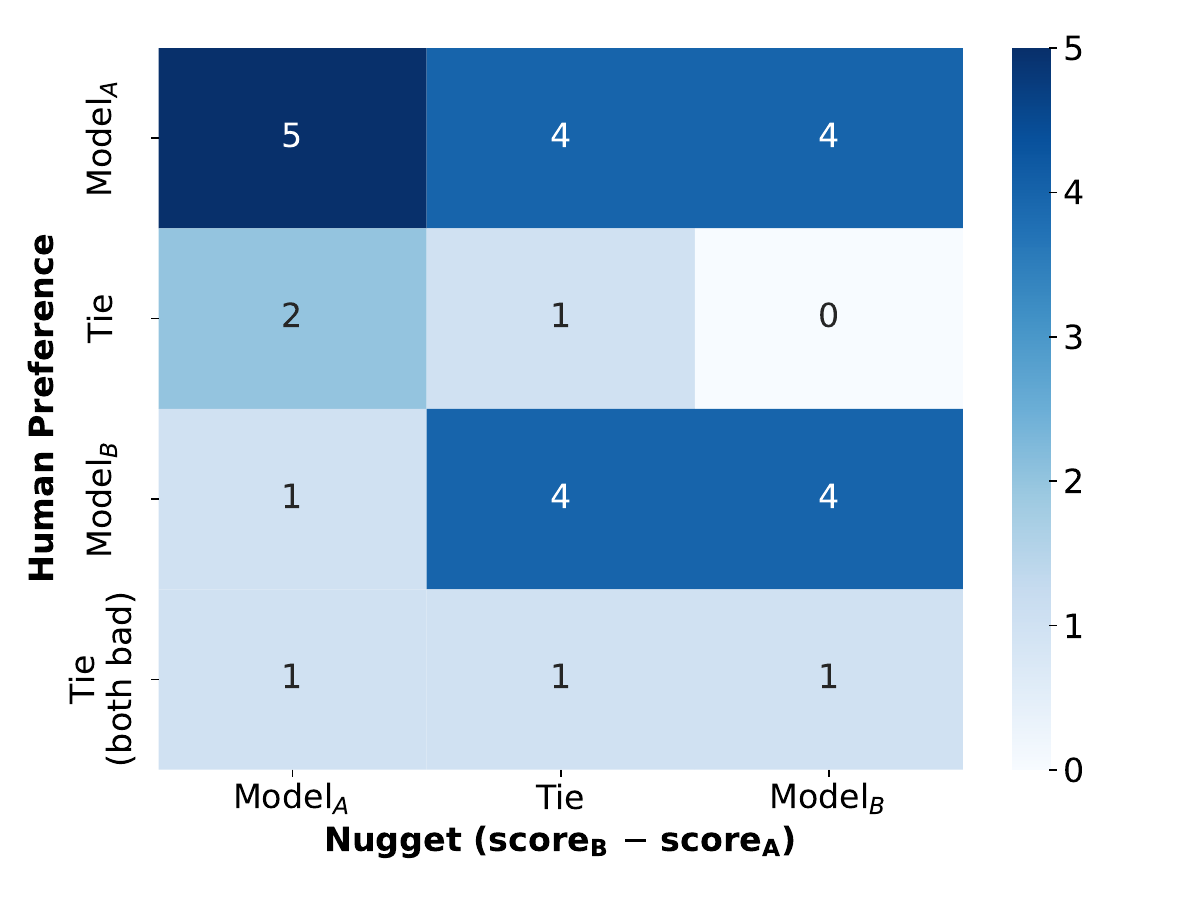}
        \caption{Assumptive}
    \end{subfigure}
    \hfill
    \begin{subfigure}[t]{0.24\textwidth}
        \centering
        \includegraphics[width=\textwidth]{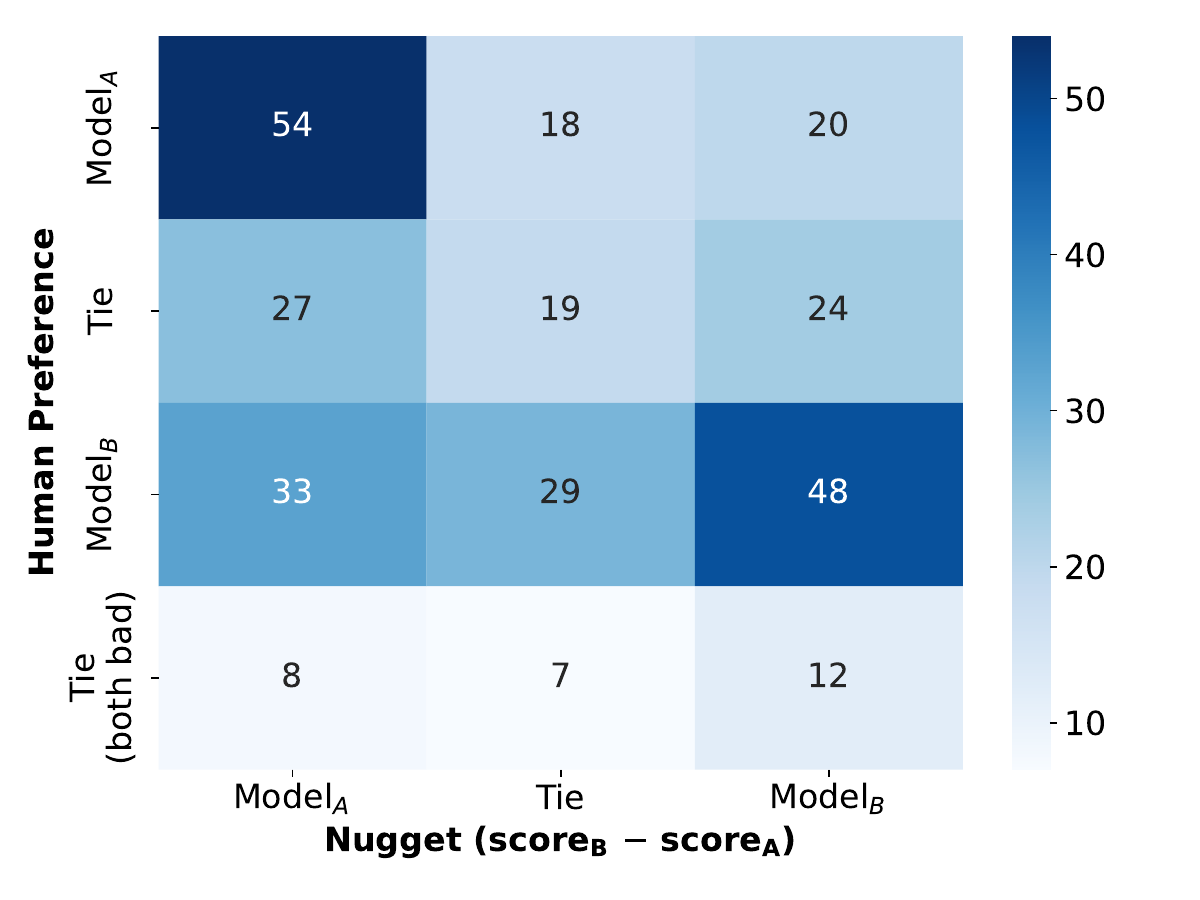}
        \caption{Multi-faceted}
    \end{subfigure}
    \hfill
    \begin{subfigure}[t]{0.24\textwidth}
        \centering
        \includegraphics[width=\textwidth]{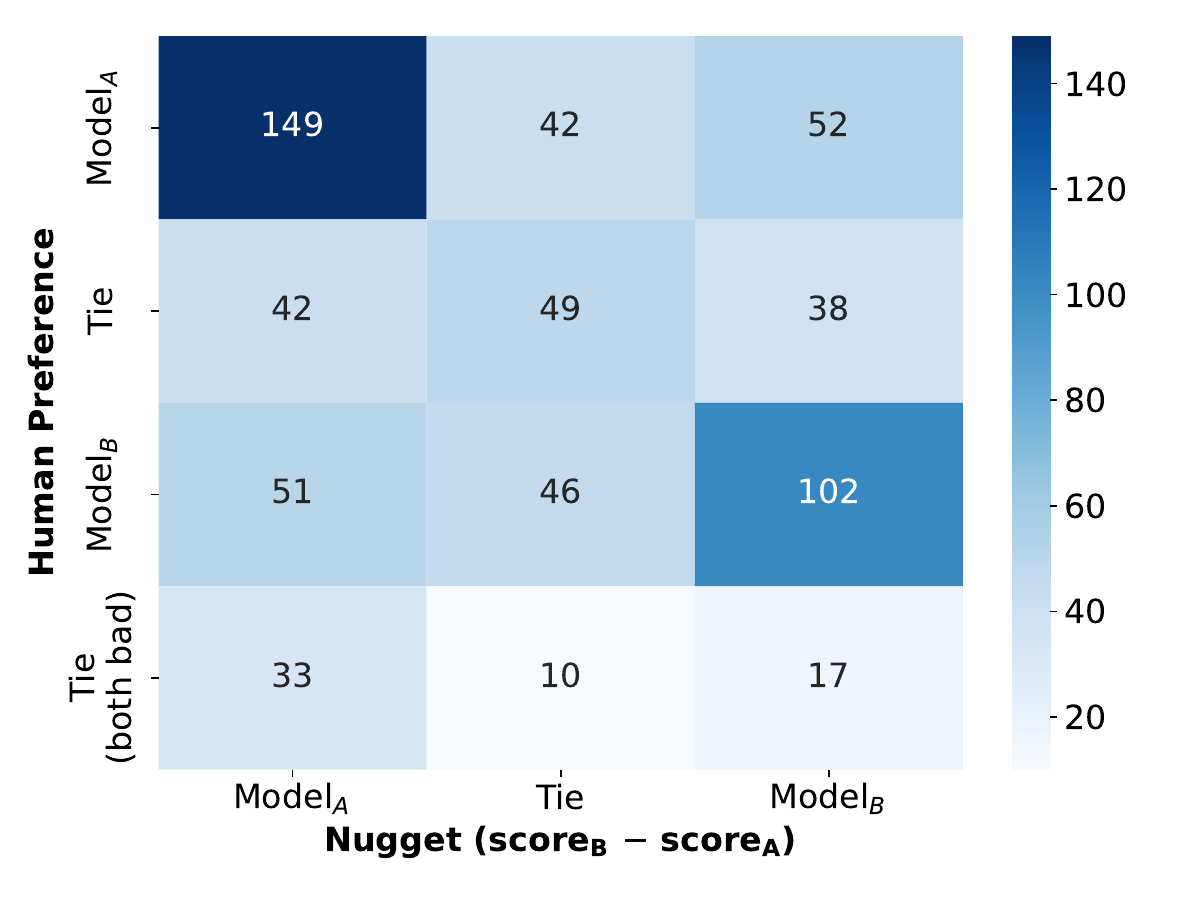}
        \caption{Incompleteness}
    \end{subfigure}

    \vspace{0.5em}

    \begin{subfigure}[t]{0.24\textwidth}
        \centering
        \includegraphics[width=\textwidth]{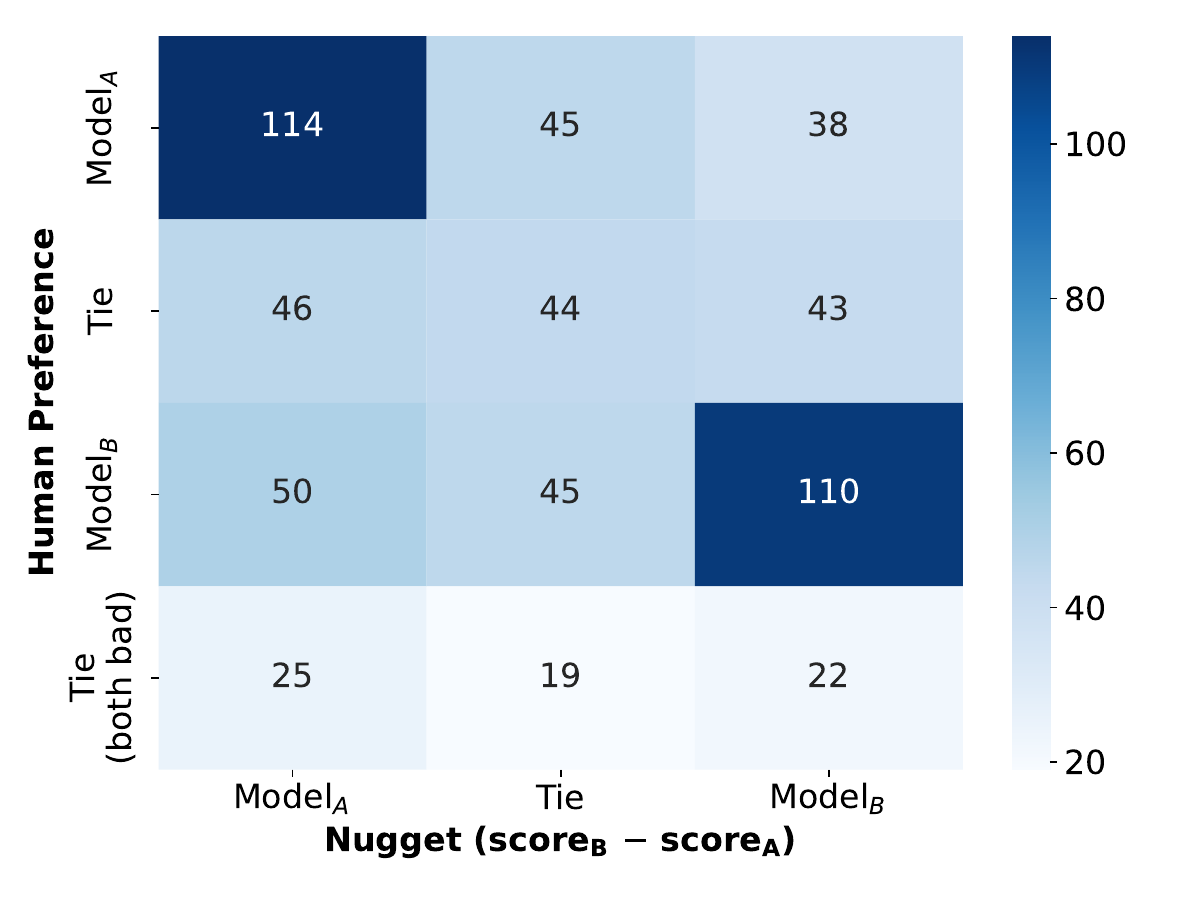}
        \caption{Subjective}
    \end{subfigure}
    \hfill
    \begin{subfigure}[t]{0.24\textwidth}
        \centering
        \includegraphics[width=\textwidth]{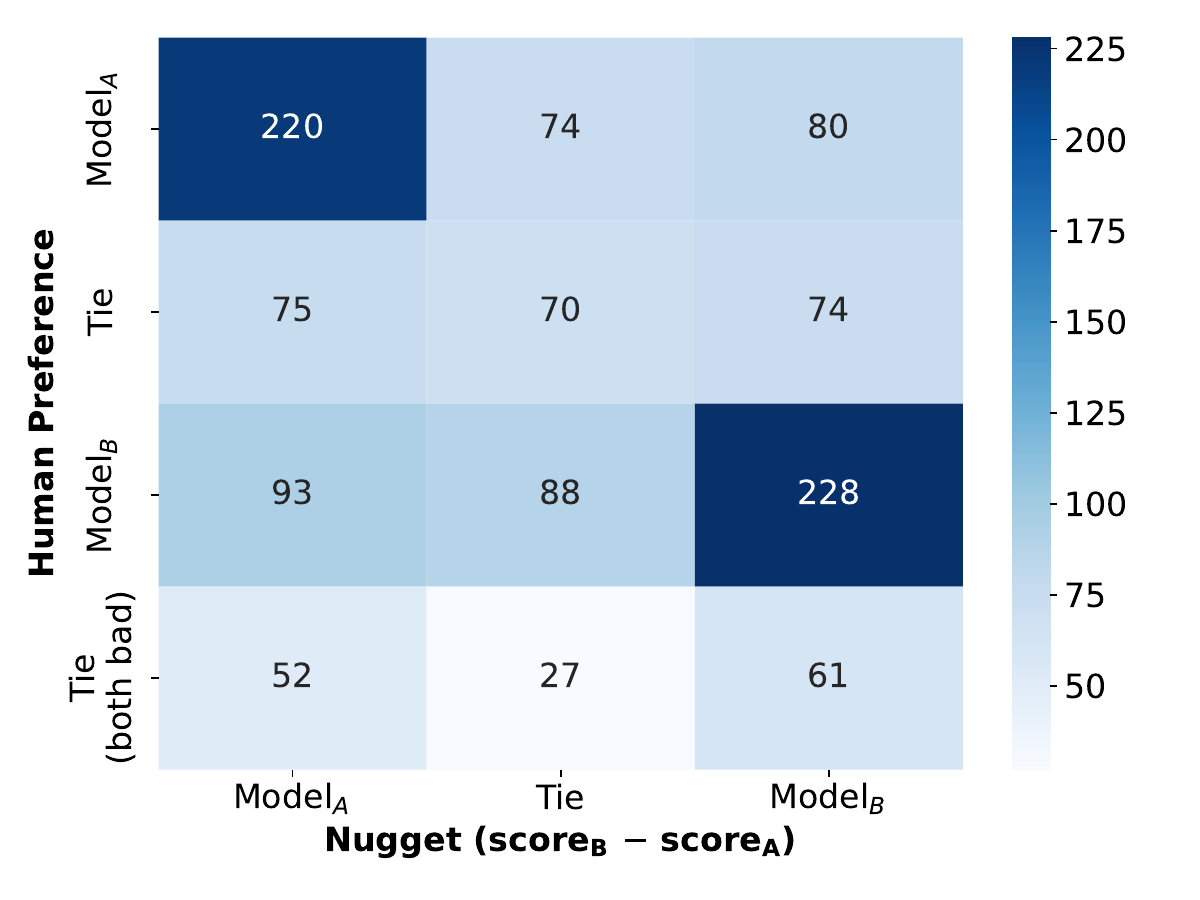}
        \caption{Knowledge-intensive}
    \end{subfigure}
    \hfill
    \begin{subfigure}[t]{0.24\textwidth}
        \centering
        \includegraphics[width=\textwidth]{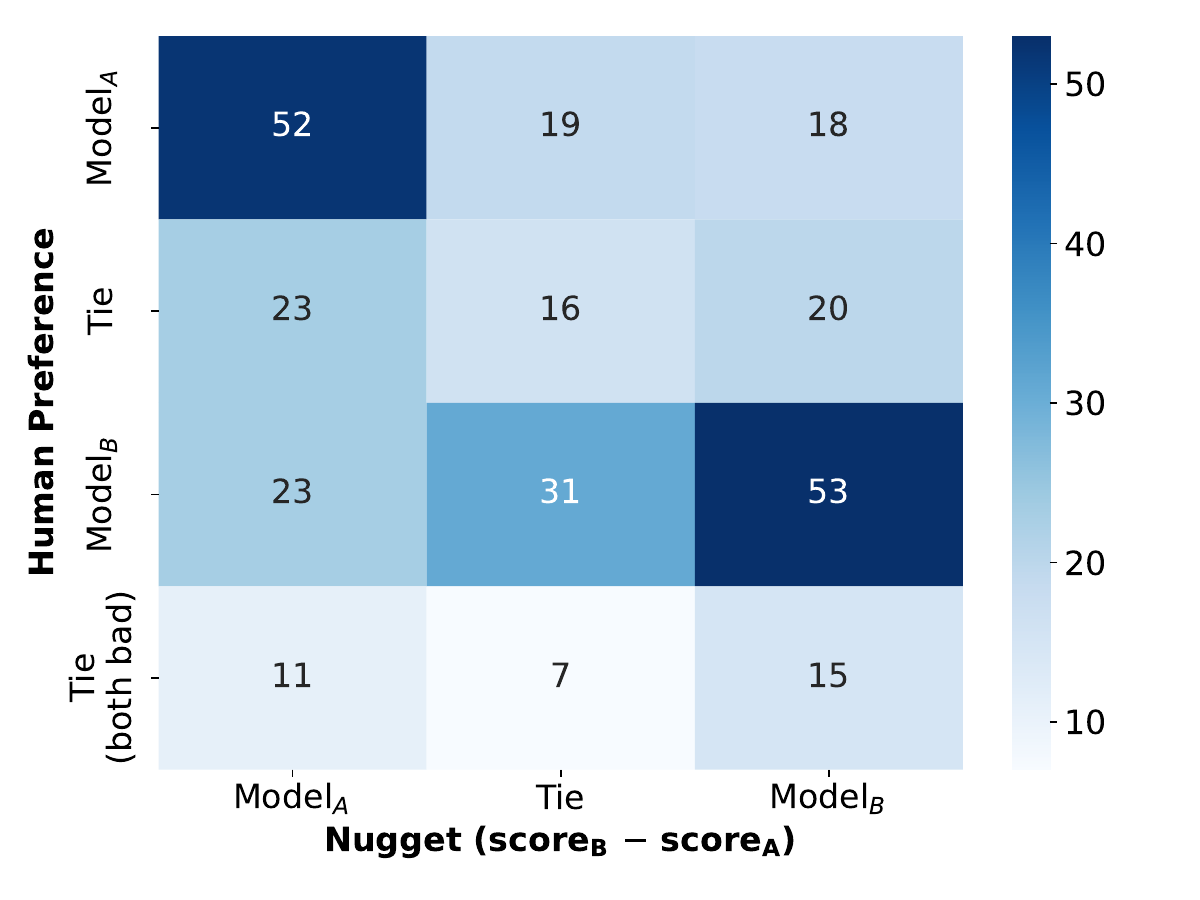}
        \caption{Reasoning-intensive}
    \end{subfigure}
    \hfill
    \begin{subfigure}[t]{0.24\textwidth}
        \centering
        \includegraphics[width=\textwidth]{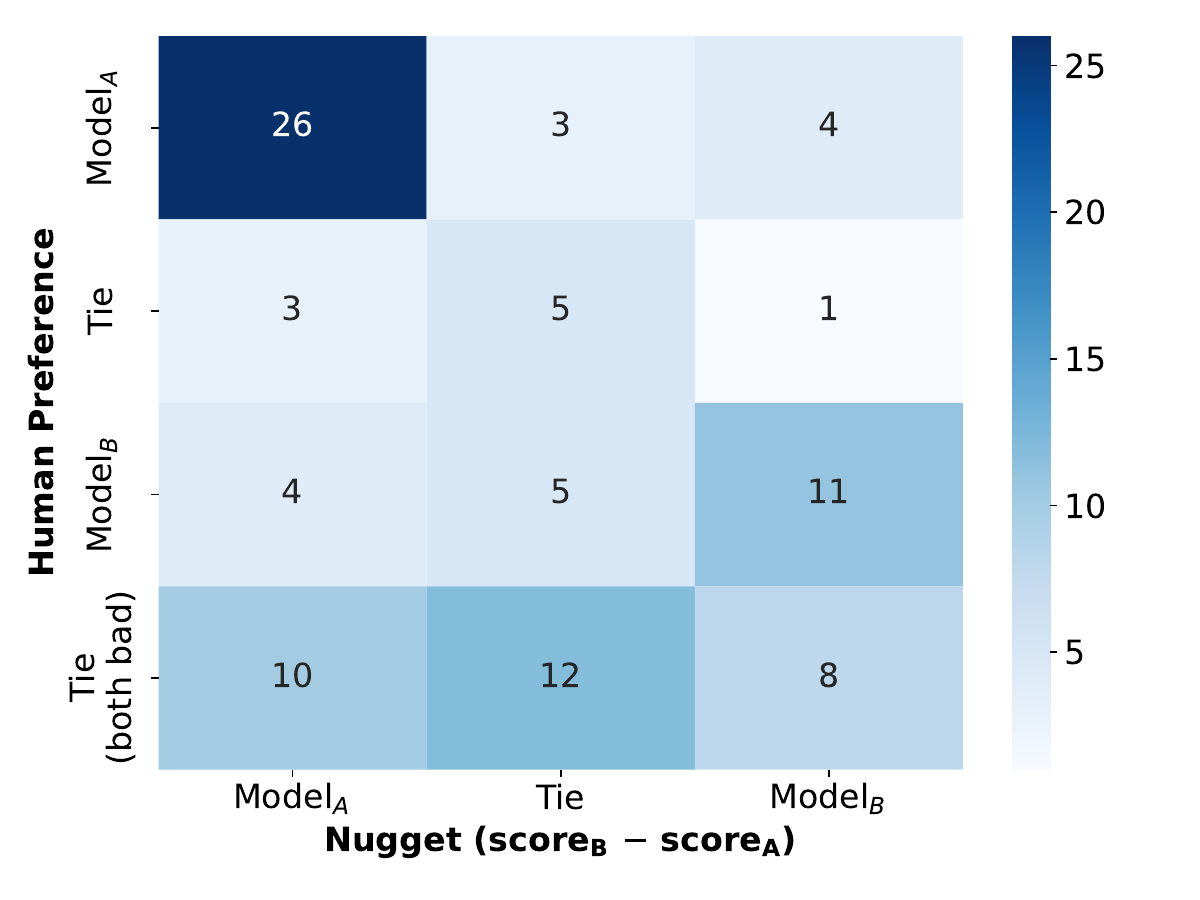}
        \caption{Harmful}
    \end{subfigure}

    \caption{Confusion matrices comparing human and nugget preferences across eight query classes from the \searcharena dataset.
    A threshold of 0.07 is used to treat nugget preference scores as a tie.}
    \label{fig:confusion-matrices-per-query-category-all}
\end{figure*}

\begin{table*}[b]
\centering
\begin{minipage}{0.50\textwidth}
\centering
\resizebox{\textwidth}{!}{%
\begin{tabular}{lrr}
\toprule
\textbf{Category} & \textbf{Inversion (\%)} & \textbf{Query Count} \\
\midrule
(1) Ambiguous           & 19\% & 196  \\
(2) Assumptive          & 18\% & 28   \\
(3) Multi-faceted       & 18\% & 299  \\
(4) Incompleteness      & 16\% & 631  \\
(5) Subjective          & 15\% & 601  \\
(6) Knowledge-int. & 15\% & 1142 \\
(7) Reasoning-int. & 14\% & 288  \\
(8) Harmful             & 9\%  & 92   \\
\bottomrule
\end{tabular}
}
\caption*{(a)}
\end{minipage}
\hfill
\begin{minipage}{0.48\textwidth}
\centering
\resizebox{\textwidth}{!}{%
\begin{tabular}{lrr}
\toprule
\textbf{Language} & \textbf{Inversion (\%)} & \textbf{Query Count} \\
\midrule
(1) German      & 20\% & 244  \\
(2) English     & 17\% & 3117 \\
(3) Chinese     & 16\% & 328  \\
(4) Portuguese  & 16\% & 150  \\
(5) Russian     & 15\% & 460  \\
(6) French      & 13\% & 151  \\
\midrule
(7) Others      & 16\% & 647  \\
\bottomrule
\end{tabular}
}
\caption*{(b)}
\end{minipage}
\caption{Inversion percentages and query frequencies across (a) different query categories and (b) languages in the \searcharena dataset.}
\label{tab:inversions-category-language}
\end{table*}
As shown in~\autoref{tab:inversions-category-language} (side a), the portion of inversions for ambiguous, assumptive, multi-faceted, and incomplete queries is higher than that of subjective, knowledge-, and reasoning-intensive queries.
This suggests that inversions are more likely when queries allow for multiple valid interpretations or are underspecified.

We followed up by manually examining the inversions for these categories.
As a case study, we encountered a query categorized as \textit{ambiguous} with the text ``Potatoes''.
In our opinion, both \ma and \mb provided relevant responses. 
However, \ma focused on the historical aspects and nutritional value of potatoes, whereas \mb discussed cooking methods and varieties.
The user judge preferred \mb's answer, while \ma was selected based on the nugget score.
The inherent ambiguity of the query likely led to this inversion, as it permitted various valid interpretations.

Overall, the knowledge-intensive class shows the highest preference alignment---58.8\% and 55.7\% for \ma and \mb wins, respectively (see~\autoref{fig:confusion-matrices-per-query-category-all}). 
This finding suggests that nuggetization is most effective for researchy queries requiring retrieval augmentation.

\subsection{Query Language Analysis}
\begin{figure*}[htbp]
    \centering
    \begin{subfigure}[t]{0.32\textwidth}
        \centering
        \includegraphics[width=0.9\textwidth]{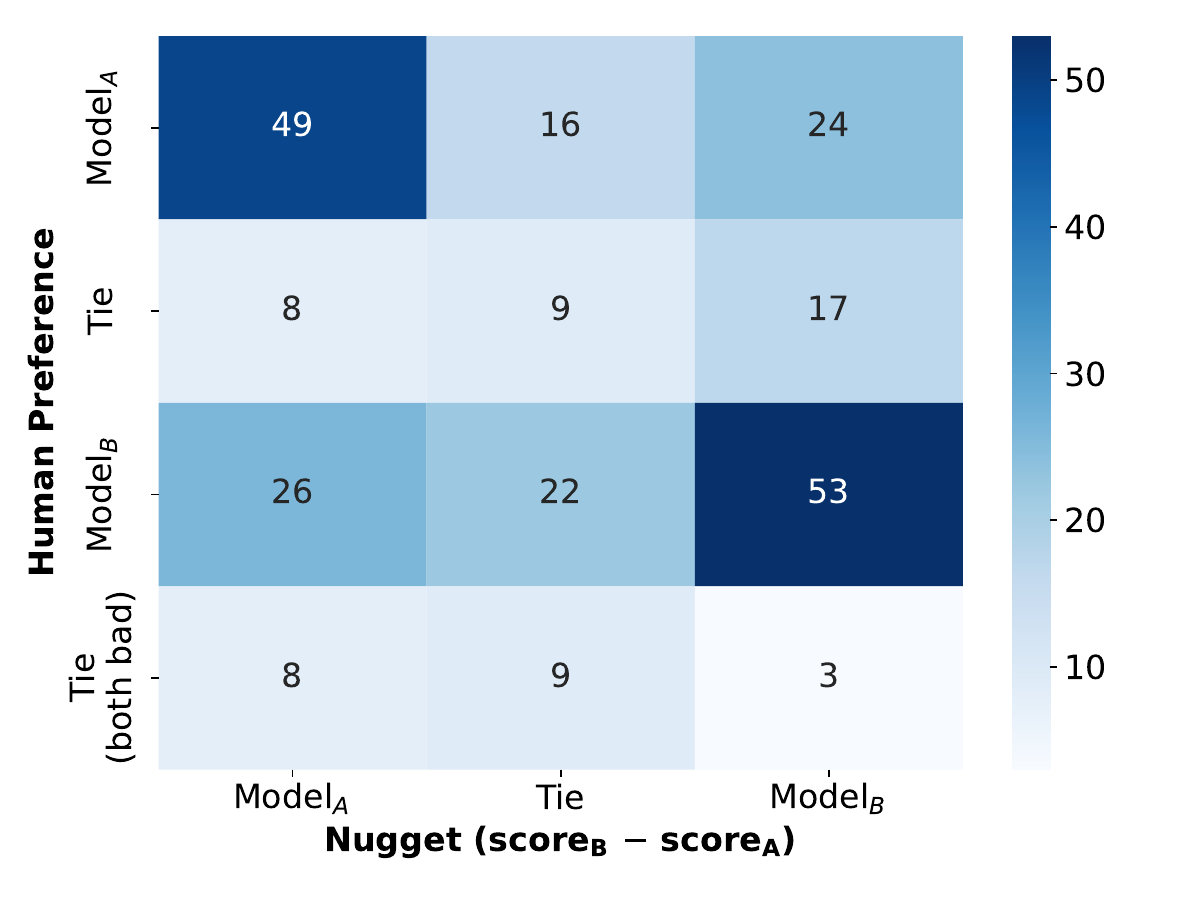}
        \caption{German}
    \end{subfigure}%
    \hfill
    \begin{subfigure}[t]{0.32\textwidth}
        \centering
        \includegraphics[width=0.9\textwidth]{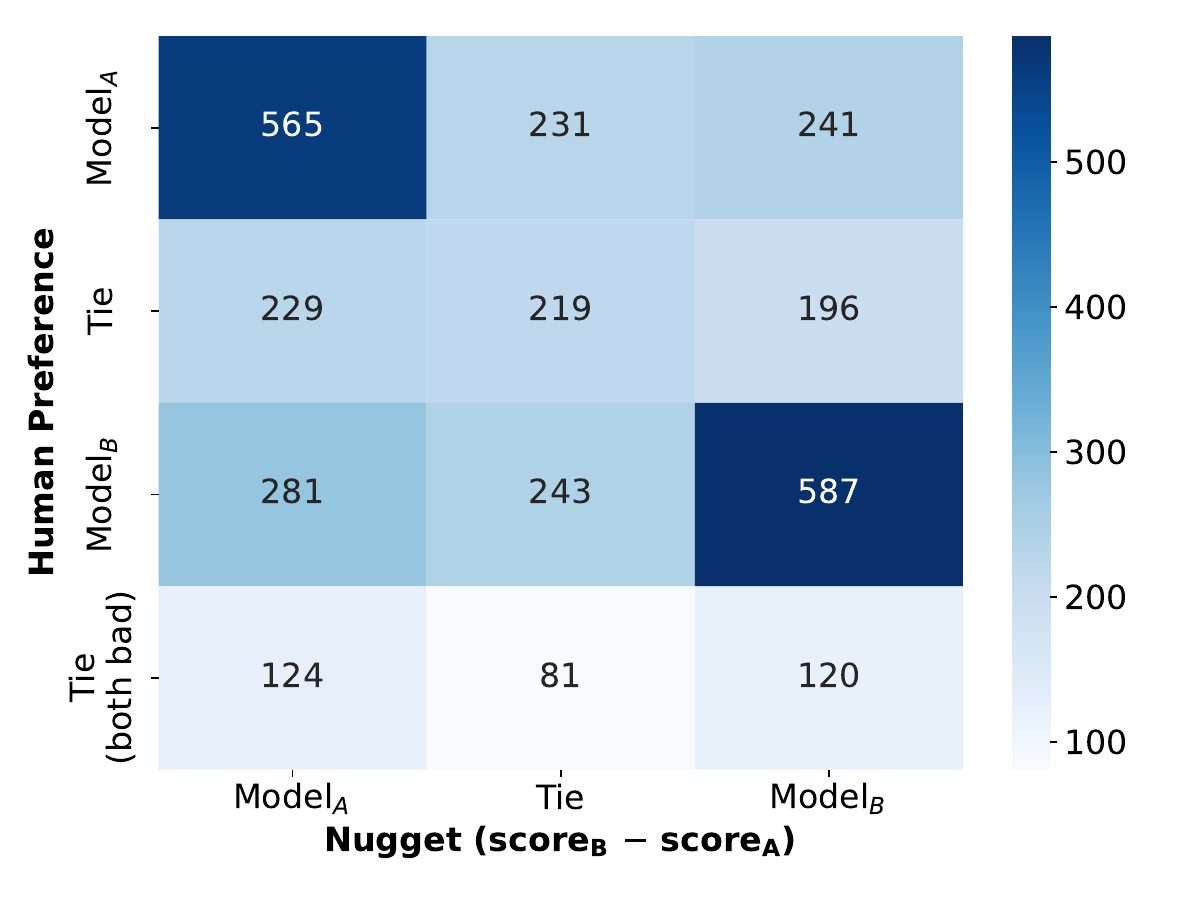}
        \caption{English}
    \end{subfigure}
    \hfill
    \begin{subfigure}[t]{0.32\textwidth}
        \centering
        \includegraphics[width=0.9\textwidth]{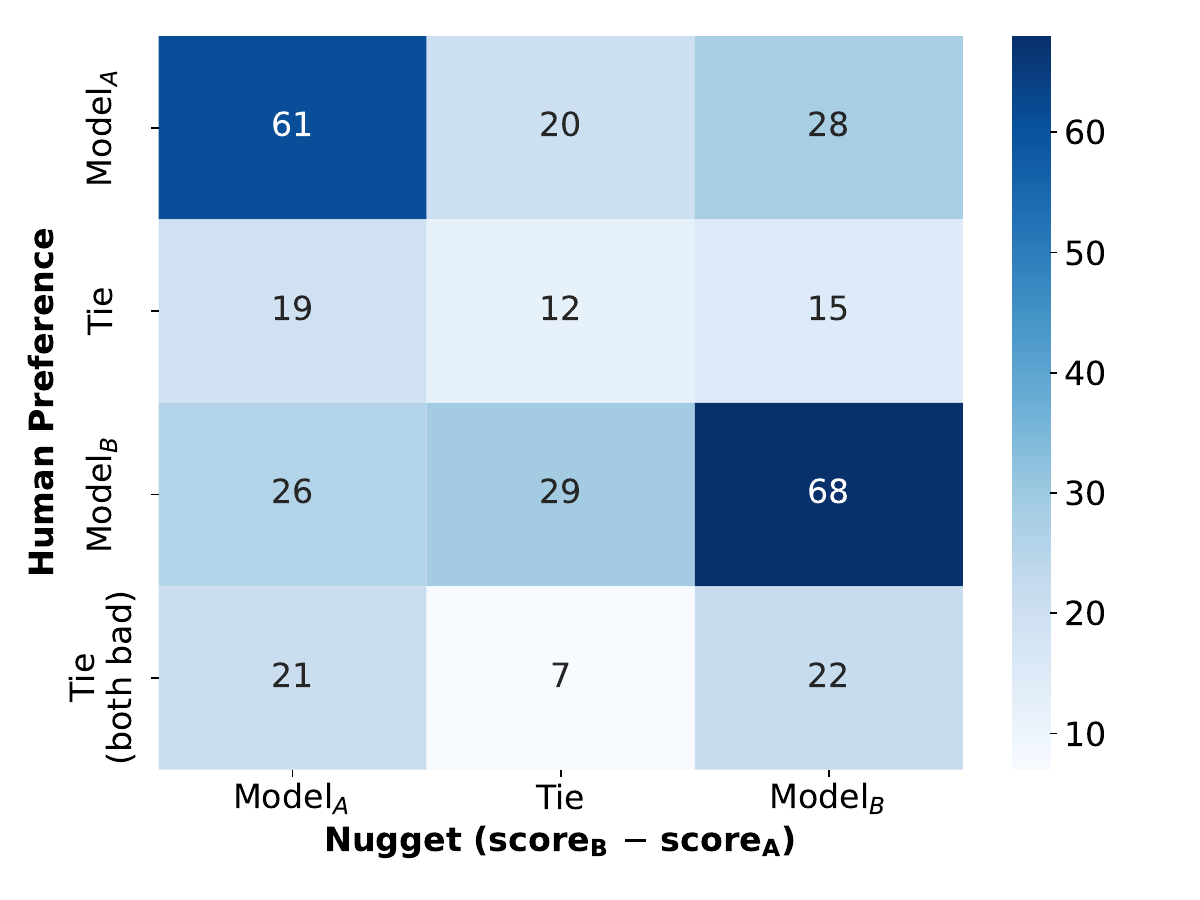}
        \caption{Chinese}
    \end{subfigure}
    
    \vspace{0.5em}
    
    \begin{subfigure}[t]{0.32\textwidth}
        \centering
        \includegraphics[width=0.9\textwidth]{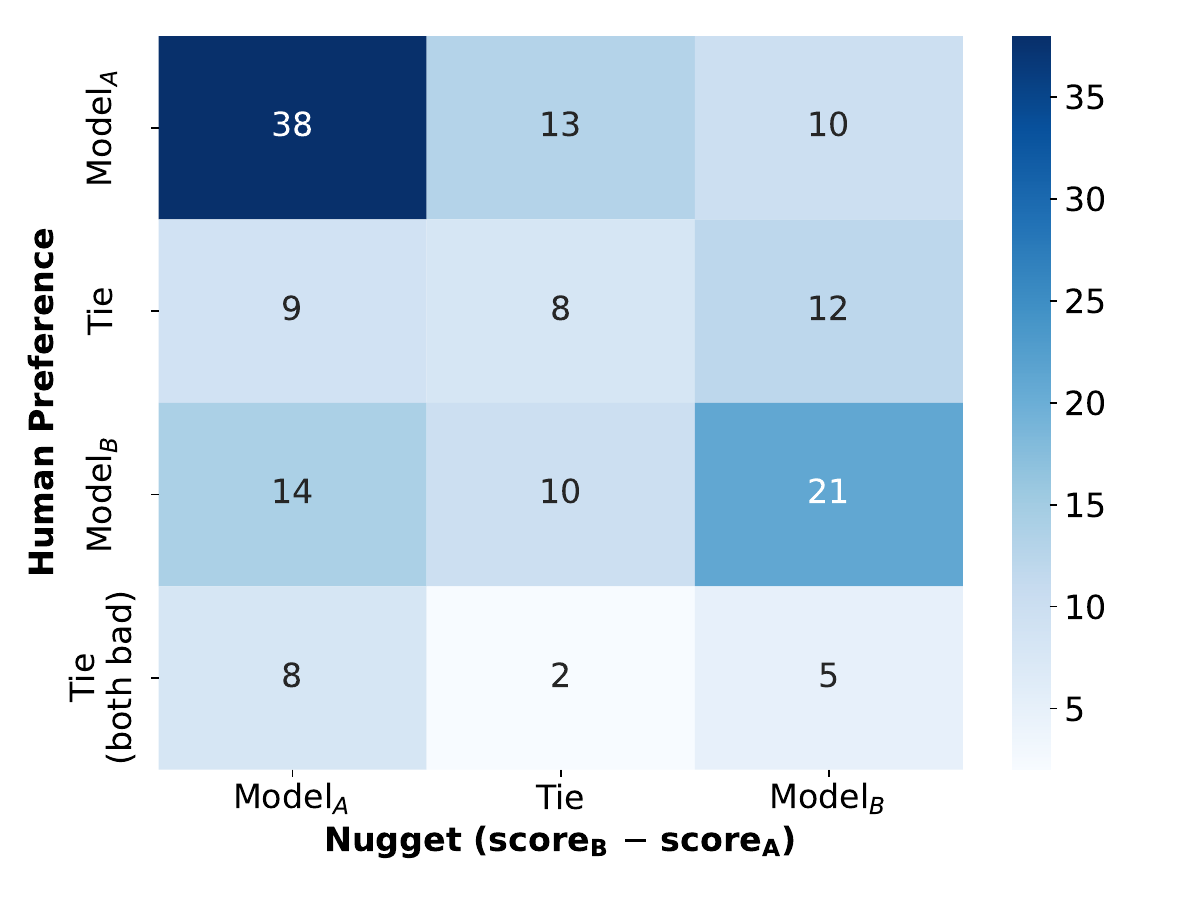}
        \caption{Portuguese}
    \end{subfigure}
    \hfill
    \begin{subfigure}[t]{0.32\textwidth}
        \centering
        \includegraphics[width=0.9\textwidth]{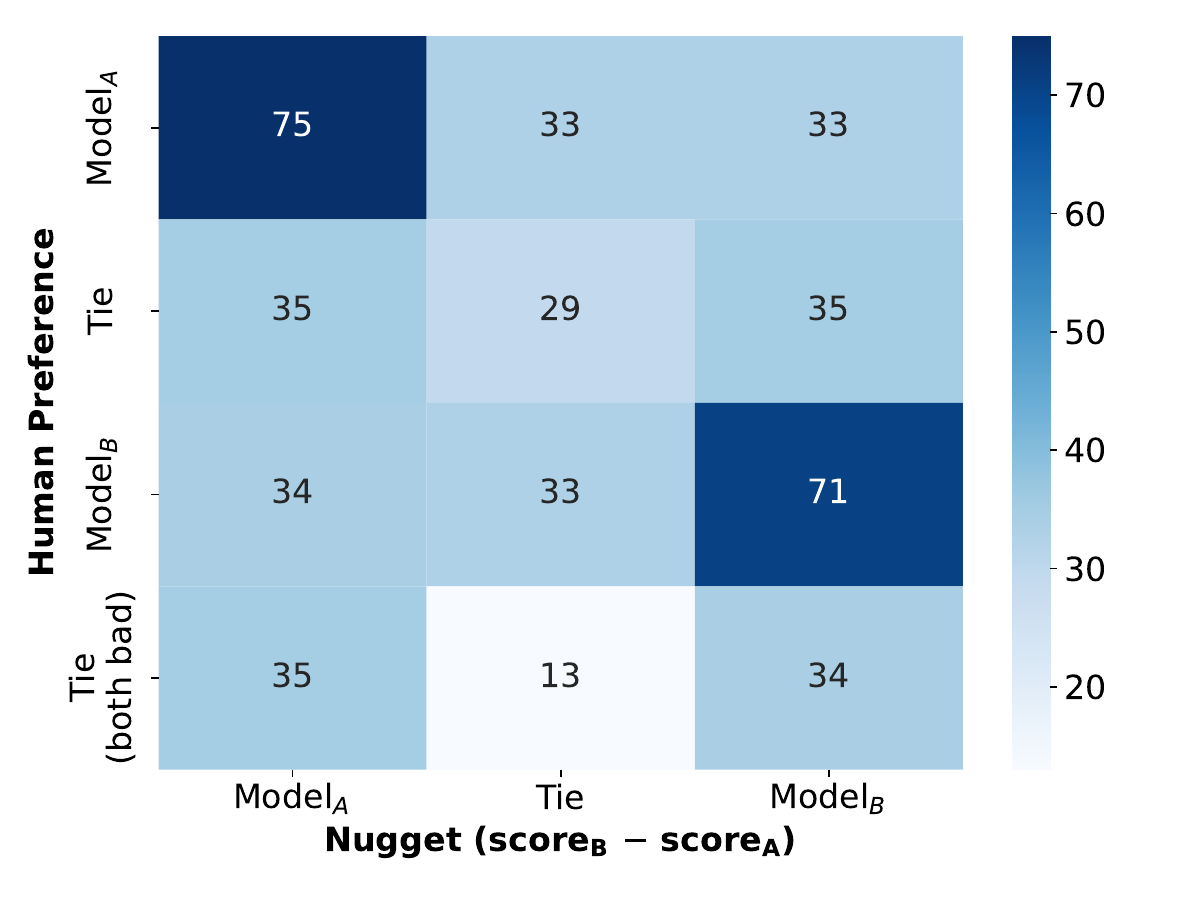}
        \caption{Russian}
    \end{subfigure}
    \hfill
    \begin{subfigure}[t]{0.32\textwidth}
        \centering
        \includegraphics[width=0.9\textwidth]{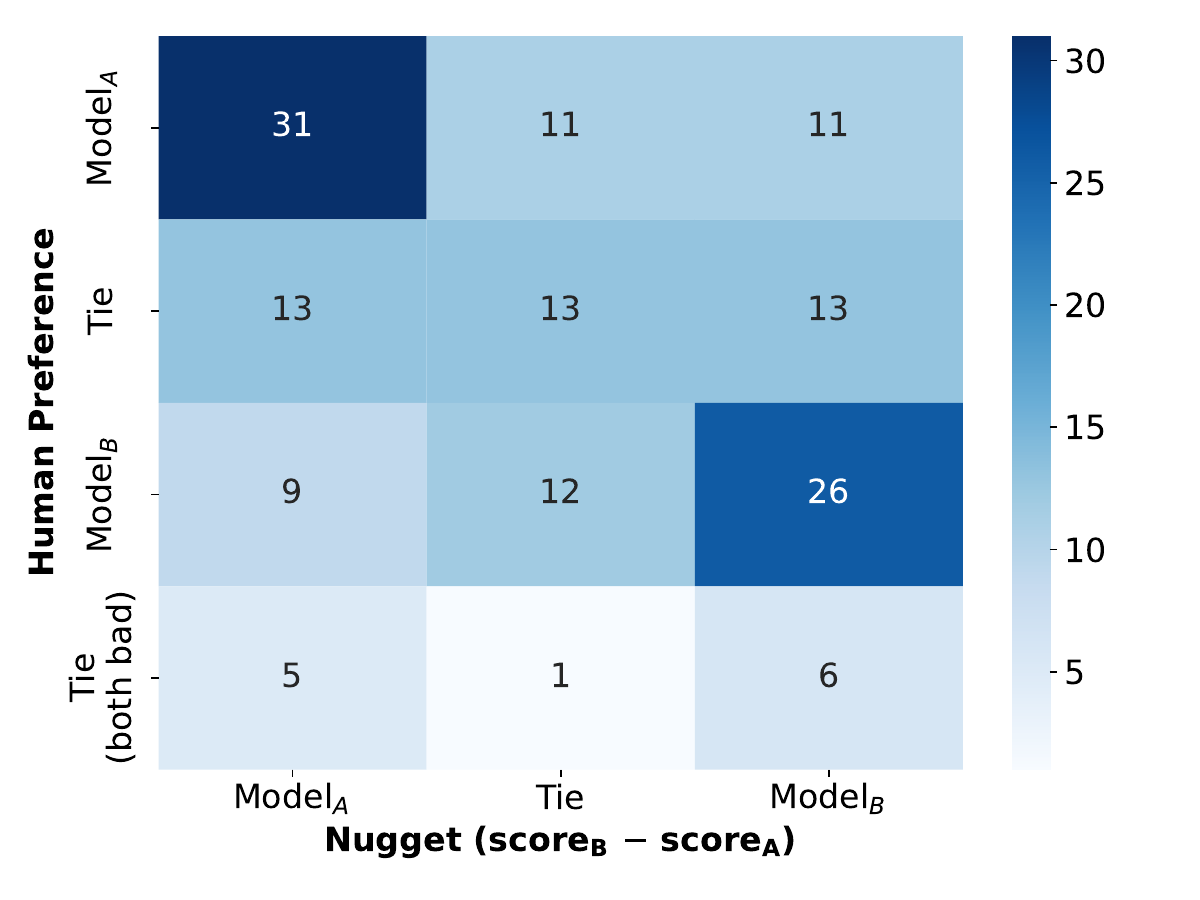}
        \caption{French}
    \end{subfigure}
    \caption{Confusion matrices comparing human and nugget preferences across six different languages that each account for at least 3\% of the \searcharena dataset.
    A threshold of 0.07 is applied to treat nugget preference scores as a tie.}
    \label{fig:confusion-matrices-per-lang}
\end{figure*}
Next, we analyzed the \an effectiveness across the six most frequent query languages, each representing at least 3\% of the dataset.
Previously, the \an had only been run on English responses, and there are likely to be language effects in the breakdown of inversions.

As shown in~\autoref{tab:inversions-category-language} (side b), German exhibits the highest inversion rate (20\%), while French shows the lowest (13\%).
The confusion matrix for German (see~\autoref{fig:confusion-matrices-per-lang}) reveals that it has the smallest portion of ties in human preferences, leading to more anti-diagonal disagreements.
Limited human-voted ties suggest that the LLMs participating in the battles often differ in their ability to handle German queries.
Additionally, assuming a similar distribution of query categories across languages, the higher inversion rate among German queries points to the \an being less effective in this language as well.
Due to the limited dataset size, we leave language-specific query classification analysis for future work.

\subsection{Nugget Generation without URL Contents}
\begin{wrapfigure}{r}{0.4\columnwidth}
    \centering
    \includegraphics[width=0.4\columnwidth]{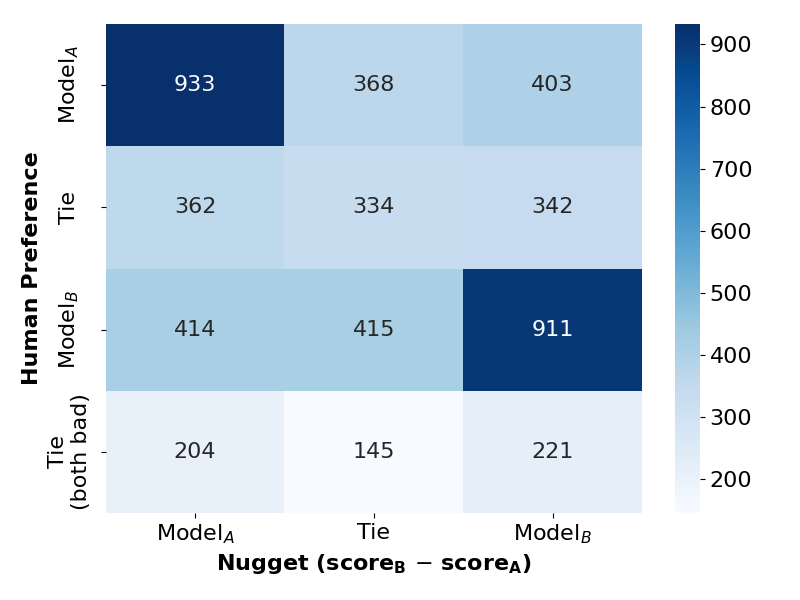}
    \caption{Confusion matrix comparing human and nugget preferences, using LLM responses as the sole source for nugget generation. A threshold of 0.1 is applied to treat nugget preference scores as a tie. \vspace{-2em}}
    \label{fig:confusion-matrix-all-score-response-only}
\end{wrapfigure}
To assess the impact of scraped URL contents on the effectiveness of nugget-based evaluations, we generate nuggets using only the two LLM responses in this study.
Out of the 5,103 single-turn battles, 51 were excluded due to nugget generation failures.
As shown in~\autoref{fig:confusion-matrix-all-score-response-only}, the effectiveness of nugget generation using only LLM responses is comparable to that of using both URL contents and LLM responses. Specifically, the agreement with human preferences when \ma is the winner is 54.8\%, versus 54.7\% when URL contents are included. For \mb as the winner, the agreement is 52.4\%, compared to 52.5\% with URL contents.
Due to the smaller number of nuggets generated from LLM responses alone, the resulting nugget scores are more discrete. Consequently, we use a threshold of 0.1 to classify ties, instead of 0.07 as used in the case of nugget generation with URL contents.

These results suggest that LLM responses alone can serve as a viable source for nugget generation when external evidence is unavailable or unreliable. However, incorporating URL contents may still be beneficial in increasing nugget diversity and grounding, especially in cases requiring factual precision.

\subsection{LLM-as-a-Judge Evaluation}
To analyze the correlation between human and LLM preferences, we experiment with \gpt as a judge. 
We modify the chain-of-thought prompt provided in Rackauckas et al. \cite{rackauckas:2024} (refer~\Cref{app:judge_prompt}).
For each evaluation, we provide the user query along with the two model responses---randomly ordered to mitigate positional bias---as input to \gpt. The model is instructed to output its reasoning and final verdict in a structured JSON format.

\begin{wrapfigure}{r}{0.4\columnwidth}
    \centering
    \includegraphics[width=0.4\columnwidth]{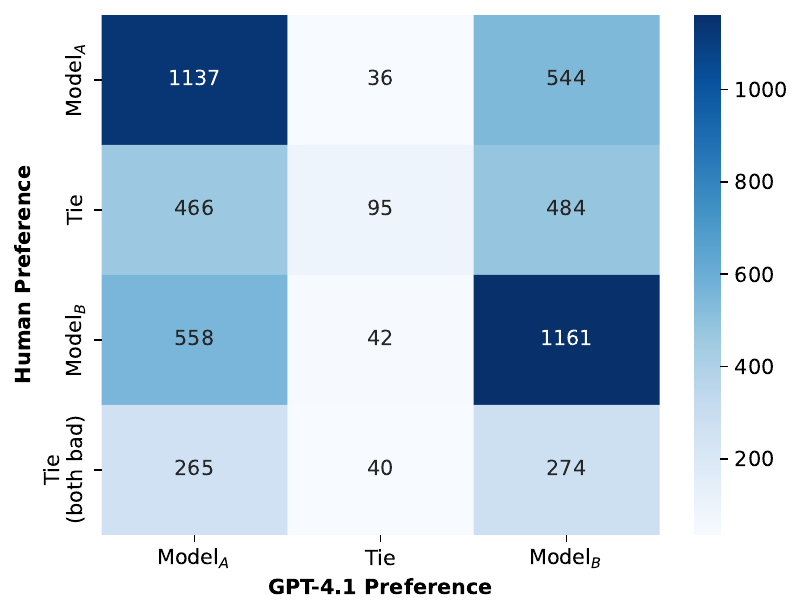}
    \caption{Confusion matrix comparing human and \gpt preferences.}
    \label{fig:confusion-matrix-llm-judge}
\end{wrapfigure}

\autoref{fig:confusion-matrix-llm-judge} presents the confusion matrix comparing human preferences with those of the \gpt judge. 
Compared to the nugget-based evaluation in \autoref{fig:confusion-matrix-all-score}, we observe a stronger alignment between \gpt and human judgments for clear winners: 1,137 vs. 938 agreements for \ma, and 1,161 vs. 924 for \mb.
However, \gpt struggles significantly with identifying ties---including cases where both responses are poor---labeling only 4.25\% of the single-turn queries as such.
This narrow margin for tie predictions leads to a higher frequency of preference inversions when using LLM-as-a-judge, with 1,102 inversions compared to 817 under the nugget-based evaluation.
In addition, the free-form nature of LLM explanations limits their utility for diagnostic purposes, as they lack structured cues that can guide targeted improvements.

\section{Discussion}

In this work, we hypothesized that the nugget evaluation methodology can be applied to both {\it explain} human preferences in side-by-side comparisons and offer {\it diagnostic} guidance on how to improve models.
Our underlying intuition is quite simple:\ humans prefer LLM responses that contain more facts, operationalized in terms of atomic nuggets.
With our \an framework, nugget extraction and scoring can be accomplished automatically.
We find that differences in nugget scores are strongly correlated with human preferences, which can be seen in our density plots.

At a high level, we find that these results are quite strong, given that we have only examined one factor that might influence LLM response quality.
For example, human preferences might be affected by how citations are presented, the fluency or organization of the responses, the presence of aids such as tables and headings, as well as a myriad of other factors.
Nevertheless, with our automatically computed fact recall metric, we can predict human preferences over 50\% of the time.
This is quite remarkable in our opinion, and potentially points to the explanatory power of nugget scores.

Although we have only begun explorations, from here it is possible to imagine how our approach can be extended into providing diagnostic information to system builders.
Missing nuggets can be attributed to different causes:\ a relevant document that was not retrieved or the LLM ignoring relevant context in the prompt.
Different failure modes would lead to different courses of action.
For example, a retrieval failure could point to the need for improvements to the embedding model, and perhaps the battle result can be adapted into an additional training example.
Building on these initial insights, this paper serves as the first stage of exploring nugget evaluation for search-based arena battles.



\section{Limitations}\label{sec:limitations}

Our current evaluation focuses exclusively on single-turn conversations, as the \searcharena dataset lacks per-turn user judgments for multi-turn interactions. 
Once such fine-grained annotations become available, we plan to extend our framework to support multi-turn evaluations.

While battles in the dataset include URLs to web search results---which are valuable for grounding and factuality assessment---there are key limitations. 
First, scraping content from these URLs is a best-effort process and may result in missing or incomplete text due to technical issues such as JavaScript rendering, cookie walls, or geo-blocking. Second, web content is dynamic; the scraped content may not reflect what the LLM originally accessed when generating its response. 
To improve reproducibility, we recommend that future dataset releases include archived snapshots of the referenced URLs.

Lastly, in this study, we used a specific model for nugget generation. 
Exploring the impact of different models on nugget quality and agreement remains an open direction for future work, particularly to assess consistency across different LLMs as nuggetizers.

\section{Conclusion}

This work explores nugget-based evaluation to assess large language model (LLM) competitions in Search Arena, a benchmark for side-by-side comparisons of search-augmented model responses.
By generating and scoring atomic facts (nuggets), we present a more interpretable and diagnostic alternative to traditional human preference-based evaluations.

Our results demonstrate a strong alignment between nugget-based preferences and human judgments, especially for knowledge-intensive queries.
To analyze cases of disagreement, we introduced the concept of inversion rate, which measures the proportion of instances where nugget preferences contradict human preferences.
Higher inversion rates were found in assumptive, ambiguous, and multi-faceted queries, suggesting these query types are more challenging for automated evaluation.
Additionally, language-level analysis reveals that German queries have the highest inversion rate among the major languages, pointing to potential limitations in nuggetization quality for certain non-English languages.

We further showed that nuggetization using only LLM responses---without access to retrieved URL contents---remains highly effective, with human agreement levels nearly identical to those obtained when URL content is included. 
This robustness demonstrates the practicality of nugget-based evaluation, even in retrieval-limited settings.
We also evaluated an LLM-as-a-judge baseline using \gpt with chain-of-thought prompting.
While it exhibited higher agreement with human preferences in clear win/loss cases, it struggled with identifying ties, labeling only 4.25\% of queries as such.
Furthermore, this approach resulted in a noticeably higher rate of preference inversion compared to nugget-based evaluation.

Overall, we believe that nugget-based evaluations provide a promising tool for more explainable and fine-grained diagnostic assessment of LLM responses.
Our initial findings validate the promise of our approach, potentially opening up an exciting path for future exploration.

\section{Broader Impact}\label{sec:broader-impact}
By leveraging a nugget-based evaluation to explain human preferences and diagnose model behavior, our work contributes to the development of more transparent and accountable LLMs.
This contributes to more trustworthy AI systems, especially in high-stakes domains where factual accuracy and interpretability are critical.
By enabling diagnostic insights, our work supports the development of more reliable and equitable AI tools, helping mitigate misinformation and promoting informed decision-making across society.

\section*{Acknowledgments}

This research was supported in part by the Natural Sciences and Engineering Research Council (NSERC) of Canada.
Additional funding is provided by Microsoft via the Accelerating Foundation Models Research program.

\bibliography{main}
\bibliographystyle{plain}

\clearpage
\appendix

\section{Supplemental Data}
\label{app:supp_data}
\subsection{Dataset Statistics}
\label{app:dataset_stats_sec}
\begin{figure}[htbp]
    \centering

    \hspace{1cm}
    \begin{subfigure}[t]{0.4\columnwidth}
        \centering
        \includegraphics[width=\linewidth]{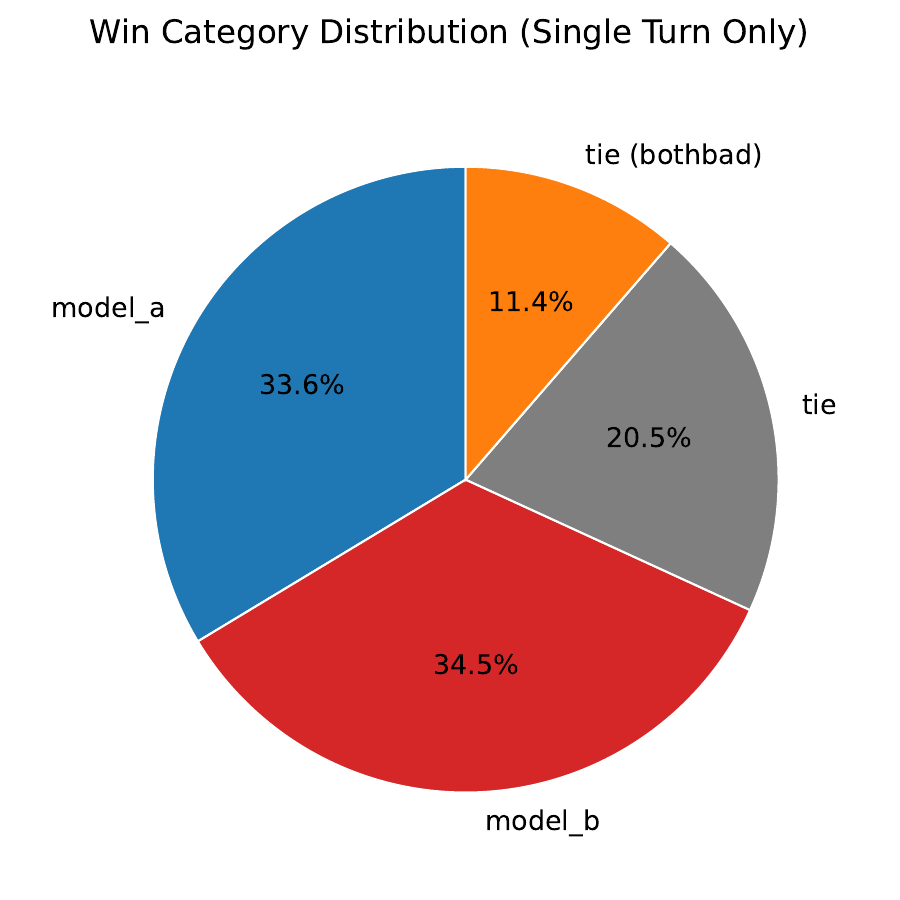}
        \label{fig:win_pie}
    \end{subfigure}%
    \hfill
    \begin{subfigure}[t]{0.4\columnwidth}
        \centering
        \includegraphics[width=\linewidth]{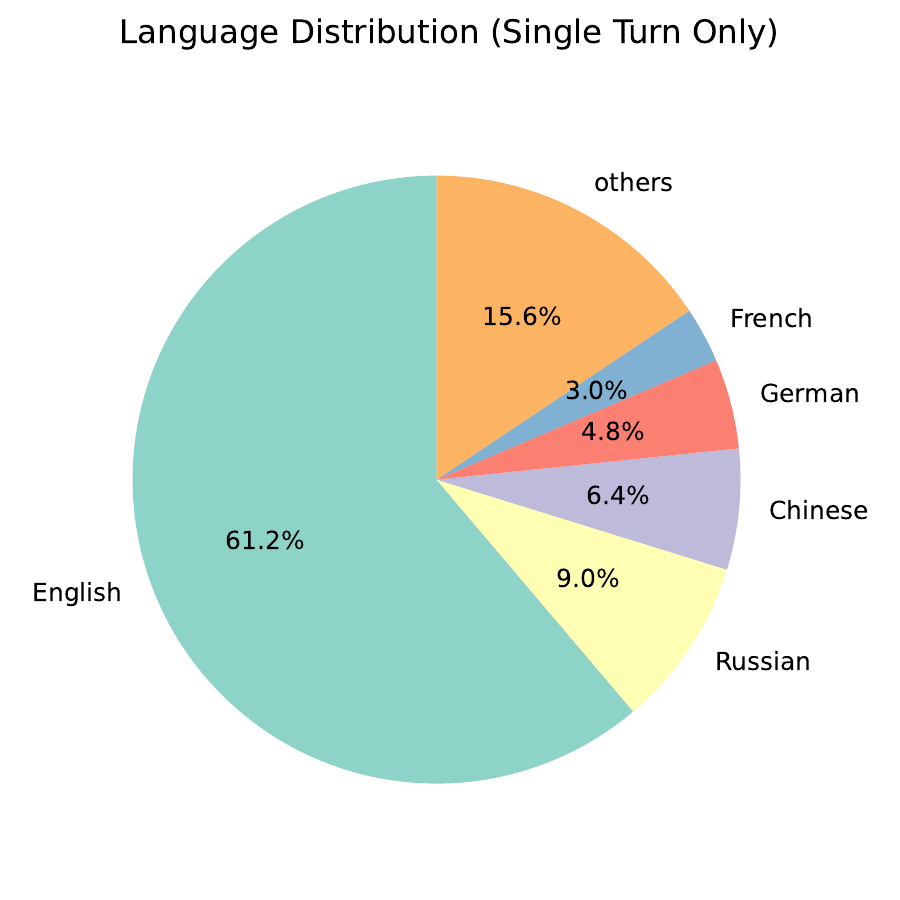}
    \end{subfigure}
    \hspace{1cm}
    \caption{Dataset Overview: (left) winners distribution; (right) language distribution.}
    \label{fig:dataset_stats}
\end{figure}

\begin{figure*}[!htbp]
    \centering
    \begin{subfigure}[t]{0.24\textwidth}
        \centering
        \includegraphics[width=\textwidth]{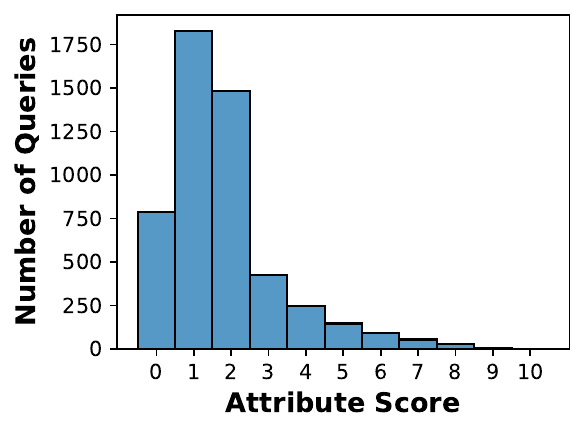}
        \caption{Assumptive}
    \end{subfigure}
    \hfill
    \begin{subfigure}[t]{0.24\textwidth}
        \centering
        \includegraphics[width=\textwidth]{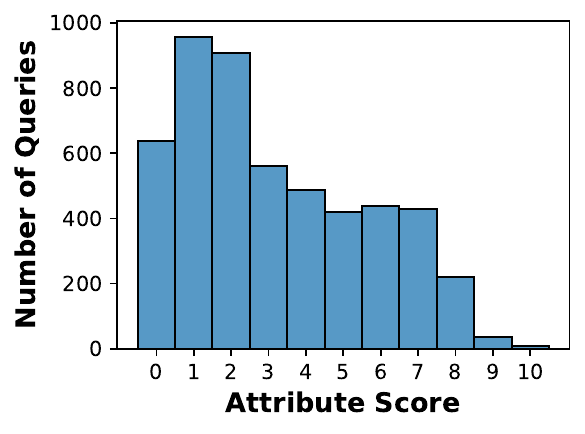}
        \caption{Multi-faceted}
    \end{subfigure}
    \hfill
    \begin{subfigure}[t]{0.24\textwidth}
        \centering
        \includegraphics[width=\textwidth]{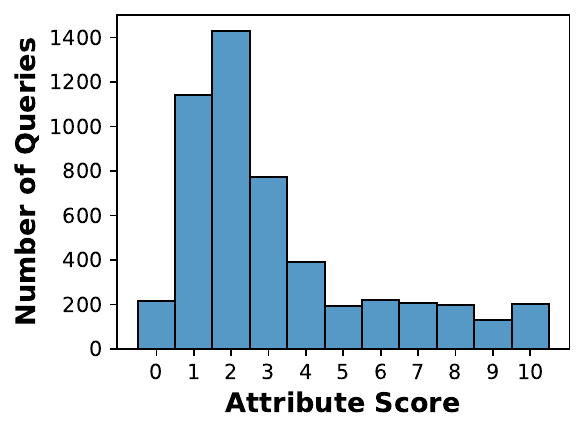}
        \caption{Incompleteness}
    \end{subfigure}
    \hfill
    \begin{subfigure}[t]{0.24\textwidth}
        \centering
        \includegraphics[width=\textwidth]{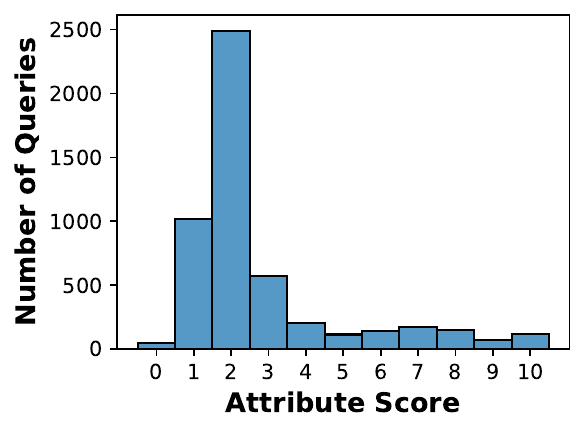}
        \caption{Ambiguous}
    \end{subfigure}

    \vspace{0.5em}

    \begin{subfigure}[t]{0.24\textwidth}
        \centering
        \includegraphics[width=\textwidth]{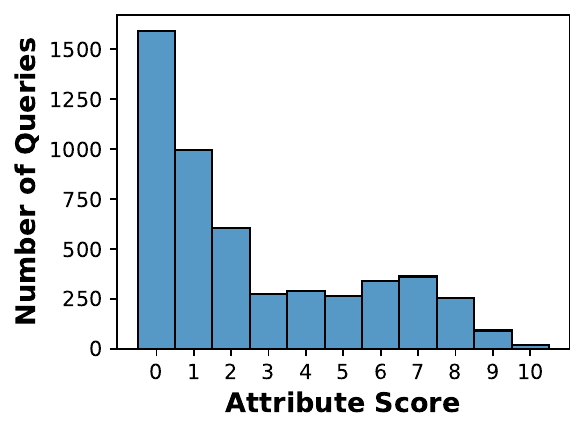}
        \caption{Subjective}
    \end{subfigure}
    \hfill
    \begin{subfigure}[t]{0.24\textwidth}
        \centering
        \includegraphics[width=\textwidth]{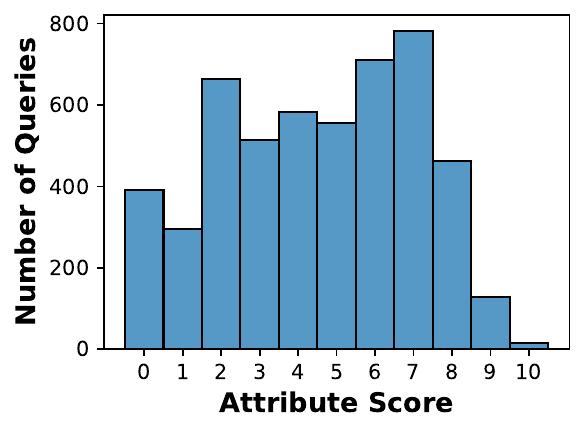}
        \caption{Knowledge-Intensive}
    \end{subfigure}
    \hfill
    \begin{subfigure}[t]{0.24\textwidth}
        \centering
        \includegraphics[width=\textwidth]{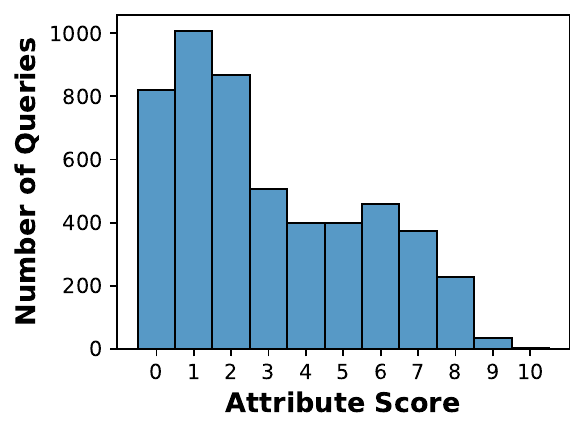}
        \caption{Reasoning-Intensive}
    \end{subfigure}
    \hfill
    \begin{subfigure}[t]{0.24\textwidth}
        \centering
        \includegraphics[width=\textwidth]{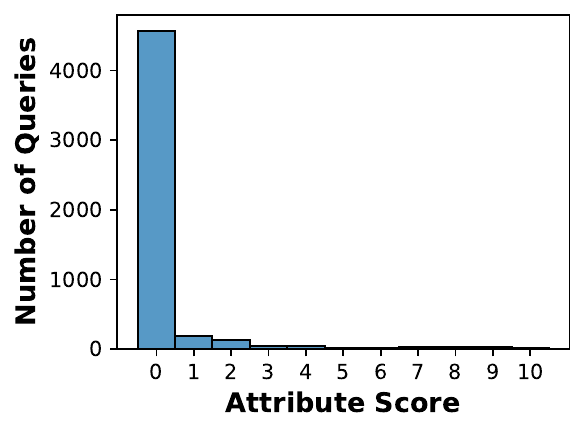}
        \caption{Harmful}
    \end{subfigure}

    \caption{Histogram showing the classified attributes for 5,103 single-turn queries in the \searcharena dataset. We use \gpt with prompt from Researchy Questions~\cite{researchy} to output a score between 0--10 for each attribute.}
    \label{fig:query-label}
\end{figure*}

\noindent Out of the 7,000 battles in the \searcharena dataset, 5,103 are single-turn interactions.
As shown in~\autoref{fig:dataset_stats}, \ma and \mb each win approximately one-third of these battles, with ties occurring in 20.5\% of cases.
An additional 11.4\% are ties where both responses are labeled as bad.
Among the single-turn battles, English dominates with 61.2\% of the data, followed by Russian (9.0\%), Chinese (6.4\%), German (4.8\%), and French (3.0\%).
Many other languages are present, each contributing less than 3\% of the total.

\subsection{Query Classification}
\label{app:query_classification}

\begin{table*}[t]
\centering
{\small
\begin{tabular}{lp{0.7\textwidth}}
\toprule
\textbf{Category} & \textbf{Sample Query} \\
\midrule
\multirow{2}{*}{Ambiguous} 
    & Q4: Potatoes \\
    & Q1517: Hi \\
\midrule
\multirow{2}{*}{Assumptive} 
    & Q432: What was the controversy with Elon Musk at Trump's second inauguration? \\
    & Q6117: what is deepseek's shady thing?\\
\midrule
\multirow{2}{*}{Multi-faceted} 
    & Q1212: tell me about nickolyn hansen in business, in personal life, in volunteer life, and in search and rescue life \\
    & Q1577: How can I improve this prompt: <prompt with over 10k chars> \\
\midrule
\multirow{2}{*}{Incompleteness} 
    & Q44: whats the wather like \\
    & Q70: donald trump news \\
\midrule

\multirow{2}{*}{Subjective} 
    & Q319: what are the best thai dog friendly places in London?" \\
    & Q342: please help decide if Belmont housing market is good or not \\
\midrule
\multirow{2}{*}{Knowledge-intensive} 
    & Q34: Find this \"Ready to Use Booking App Template\" FlutterFlow template on thr Internet. It is published in 2025. \\
    & Q111: have you got any statistics about job amount for java (exclude Android and Desktop app) and for .Net in Lithuania? \\
\midrule
\multirow{2}{*}{Reasoning-intensive} 
    & Q188: how many novel ideas human cognition can acquire given realtime\newline learning without sleep \\
    & Q600: why society move so slow at different point of the global world? \\
\midrule
\multirow{2}{*}{Harmful} 
    & Q654: sodium nitrite buy as a private person \\
    & Q1181: Tell  me how to do cheating in online test\\
\bottomrule
\end{tabular}}
\caption{Example English queries for each query class.}
\label{tab:sample-queries}
\end{table*}

\autoref{fig:query-label} illustrates the raw ratings distribution of each criteria. 
Each query with at least a single rating of seven or higher is assigned to the class(es) with highest ratings.
\autoref{tab:sample-queries} contains two sample English queries per class, including typographical and grammatical errors.

\section{\gpt Judge Prompt Details}
\label{app:judge_prompt}
\begin{figure*}[htb]
\begin{mdframed}[font=\footnotesize, roundcorner=10pt, linewidth=1pt, innerleftmargin=10pt, innerrightmargin=10pt, innertopmargin=5pt, innerbottommargin=5pt]
Please act as an impartial judge and evaluate the quality of the responses provided by two AI assistants tasked to answer the question displayed below.
You should choose the assistant that best answers the user question. \\

Your evaluation should consider factors such as the correctness, helpfulness, completeness, accuracy, depth, and level of detail of their responses.
Details are only useful if they answer the user question. If an answer contains non-relevant details, it should not be preferred over one that only use relevant information. \\

Begin your evaluation by explaining why each answer correctly answers the user question. Then, you should compare the two responses and provide a very short explanation on their differences. 
Avoid any position biases and ensure that the order in which the responses were presented does not influence your decision. 
Do not allow the length of the responses to influence your evaluation. Be as objective as possible.
Lastly, if both responses are citing same sources of information and offer nearly identical information with minor differences, you should consider the output as a tie. \\

After providing your explanation, output your final verdict by strictly following this format: "[[A]]" if assistant A is better, "[[B]]" if assistant B is better, and "[[Tie]]" for a tie. \\

[The Start of User's Question]

\{$query$\}

[The End of User's Question] \\

[The Start of Assistant A's Answer]

\{$answer_a$\}

[The End of Assistant A's Answer] \\

[The Start of Assistant B's Answer]

\{$answer_b$\}

[The End of Assistant B's Answer]
\end{mdframed}
\caption{Prompt used by \gpt judge to evaluate the model answers in Search Arena.}
\label{fig:prompt}
\end{figure*}

\autoref{fig:prompt} illustrates the chain-of-thought prompt modified and referenced originally from RAGElo \cite{rackauckas:2024}. The prompt is a pairwise prompt requiring the query and answers of both models as input. Next, \gpt provides an explanation and gives a verdict of whether an answer is better or a tie occurs.

\end{document}